\documentclass[fleqn,usenatbib]{mnras}

\usepackage{newtxtext,newtxmath}

\usepackage[T1]{fontenc}
\usepackage{ae,aecompl}

\usepackage{amsmath}
\usepackage{graphicx}
\usepackage{natbib}
\usepackage{color}
\usepackage{soul}

\newcommand{\C}{\textsc{Cloudy}}
\newcommand{\msun}{M$_{\odot}$}
\newcommand{\rsun}{R$_{\odot}$}

\title[Optical emission lines of symbiotic binaries]{Optical emission line spectra of symbiotic binaries}

\author[J. Kuuttila \& M. Gilfanov]{
J. Kuuttila$^{1}$,
M. Gilfanov$^{1,2}$,
\\
$^{1}$Max Planck Institute for Astrophysics, Karl-Schwarzschild-Str. 1, Garching b. M\"unchen 85741, Germany\\
$^{2}$Space Research Institute, Profsoyuznaya 84/32, 117997, Moscow, Russia\\
}

\date{Accepted XXX. Received YYY; in original form ZZZ}

\pubyear{2020}

\begin{document}
\label{firstpage}
\pagerange{\pageref{firstpage}--\pageref{lastpage}}
\maketitle

\begin{abstract}

Symbiotic stars are long-period interacting binaries where the compact objects, most commonly a white dwarf, is embedded in the dense stellar wind of an evolved companion star. UV and soft X-ray emission of the accretion disk and nuclear burning white dwarf plays a major role in shaping the ionisation balance of the surrounding wind material and giving rise to the rich line emission.  
In this paper, we employ 2D photoionisation calculations based on \C\ code to study the ionisation state of the circumbinary material in symbiotic systems and to predict their emission line spectra. Our simulations are parameterized via the orbital parameters of the binary and the wind mass-loss rate of the donor star, while the mass accretion rate, temperature and luminosity of the WD are computed self-consistently. We explore the parameter space of symbiotic binaries and compute luminosities of various astrophysicaly important emission lines. The line ratios are compared to the traditional diagnostic diagrams used to distinguish symbiotic binaries from other types of sources and it is shown how the binary system parameters shape these diagrams. In the significant part of the parameter space the wind material is nearly fully ionized, except for the ``shadow'' behind the donor star, thus the WD emission is typically freely escaping the system.

\end{abstract}




\section{Introduction}

Symbiotic binaries are interacting binaries consisting of an evolved giant donor star and a compact object, typically a white dwarf (WD) but it can also be a neutron star. The compact object is often accreting gas efficiently from the wind of the donor star, so that hydrogen fusion can proceed steadily on its surface giving rise to high ($\gtrsim 10^5$ K) temperatures and luminosities ($\gtrsim 10^{37}$ erg s$^{-1}$). 
In this regime, symbiotic binaries are often considered as possible progenitor candidates for Type Ia supernovae, due to their possibly high  mass accretion rates  \citep[e.g.][]{Wang18, Ilkiewicz20}.
At lower mass accretion rates, the accretion disk and boundary layer can also become significant sources of ionizing radiation. Symbiotic binaries are thus often characterised by emission lines from high ionization states, for example He~\textsc{ii}, O~\textsc{vi}, and Fe~\textsc{x}, in addition to many bright lower ionization state lines such as H Balmer lines. Symbiotic binaries are also characterised by a complex circumstellar environment resulting from the hot ionizing compact object embedded in the dense neutral wind of the donor star. Symbiotic binaries are classified into two main categories based on their near-IR spectrum: stellar continuum emission dominated S-type and dust continuum dominated D-type binaries \citep{Allen74, Webster75, Akras19}.
For a recent review of symbiotic binaries see \citet{Mikolajewska12,Munari19}.

The radii of the red giants are in the order of astronomical units, so also the orbital separations in symbiotic binaries are measured in AUs. Consequently, the orbital periods of (S-type) symbiotic binaries range mostly from 1 to 6 years \citep{Gromadzki13}, while some Mira type variables (D-type) have longer periods, up to tens of years \citep[e.g.][]{Mikolajewska09}. Thus measuring the orbital periods of symbiotic binaries can be a difficult task, which is made even more difficult by the variable nature of these binaries. 
Instead of long monitoring programs, some insight into the symbiotic binaries can be gained from the optical spectrum of the system. Especially some high ionization state lines, for example [Fe~\textsc{x}], are very sensitive to binary properties and thus detecting and measuring these lines can provide vital information about the symbiotic binaries quite easily. 

So far only $\sim$ 300 symbiotic binaries are known in the Milky Way \citep{Akras19, Merc19}, which is much lower than predicted by the population modelling simulations, which range from a few thousand up to $4 \times 10^5$ \citep{Magrini03, Lu06, Yungelson10}. The reason is not fully clear, but the possible explanations include significant absorption by circumstellar and interstellar medium, and the confusion with other types of astronomical sources. Especially often symbiotic binaries have been mistaken for planetary nebulae and dense H \textsc{ii} regions \citep{Belczynski00}, while confusion with Be stars and young stellar objects is also notable \citep{Corradi08, Rodriguez-Flores14}.
So far the most reliable features to identify the symbiotic binaries are the Raman scattered O \textsc{vi} features at $\lambda\lambda$6830 and 7088 \AA\AA, which are almost exclusively observed in the symbiotic binaries. However, these features are present only in $\approx 55\%$ of the Galactic symbiotic binaries \citep{Allen80, Schmid89, Akras19}. 

Various emission line ratio diagrams have been proposed to distinguish symbiotic binaries from planetary nebulae and other sources. \citet{Gutierrez-Moreno95} used the [O \textsc{iii}] and Balmer emission lines for the classification, but this method is not always applicable, because the [O \textsc{iii}] are not always detected. \citet{Ilkiewicz17} explored various other emission lines including [N \textsc{ii}], [O \textsc{iii}], [Ne \textsc{iii}], and He \textsc{i} lines, in order to distinguish planetary nebulae from symbiotic binaries. They demonstrated that this set of lines provides an efficient diagnostics for symbiotic binaries. It also follows from their results, that the best single  diagnostics is provided by the [Ne \textsc{iii}] thanks to its high ionization potential and high critical density.

However, there exist many symbiotic binaries with no or little forbidden line emission and thus these methods cannot always be used. 
\citet{Proga94} analysed the He \textsc{i} $\lambda \lambda$5876, 6678, 7065 emission lines and showed that these lines provide a useful diagnostic tool for the symbiotic binaries and they can be used to separate between S- and D-type symbiotics. \citet{Ilkiewicz17} showed that these lines can also be used to separate symbiotic binaries from planetary nebulae. The He \textsc{i} line ratios in symbiotic binaries differ considerably from the standard Case B approximation due to the metastability of the 2$^3$S level and collisional effects and thus these line ratios are very sensitive to the physical conditions in the highest density regions \citep{Schmid90}.

Recently \citet{Kuuttila21} studied a symbiotic binary LIN 358 with a combination of optical spectroscopy and 2D photoionization simulations and measured all main parameters of the system, including the WD temperature, the mass accretion rate and the mass-loss rate from the giant donor star. 
Here we utilise this method to explore the parameter space of symbiotic binaries with 2D photoionization simulations. We present grids of emission models for a large range of mass-loss rates and orbital separations and provide observational predictions for various emission lines as a function of the mass-loss rate and binary separation. Comparing our results with the traditional  emission line ratio diagrams used for distinguishing between symbiotic binaries and planetary nebulae, we explore how the binary system parameters determine locations of different symbiotic systems on these diagrams. 

The paper is organised as follows: in Sec.~\ref{sec:model} we introduce our model and simulations, in Sec.~\ref{sec:results} we present our results for a 0.6 \msun\ WD while the results for a 1.0 \msun\ WD are shown in the Appendix, and then we summarise our findings in Sec.~\ref{sec:conclusions}.

\section{Model description}\label{sec:model}

\subsection{Cloudy simulations}

In order to study the emission line spectra of accreting symbiotic binaries, we have run a grid of simulations with the photoionization and spectral synthesis code \C\ version 17.02 \citep{Ferland17}. The simulations were performed in a 2D configuration as introduced by \citet{Kuuttila21}. In this method the ionizing source, i.e. the white dwarf, is located at a distance $r_c$ from the centre of the spherically symmetric density distribution, i.e. the centre of the mass-losing donor star. We assume the mass-loss to be of the form \citep[e.g.][]{Nussbaumer87}
\begin{equation}\label{eq:massloss}
    \dot{M} = 4 \, \pi \, \mu \, m_{\mathrm{H}} \, v_w \, r^2 \, n(r) \, \left( 1 - \frac{R_*}{r} \right) ,
\end{equation}{}
where $\mu$ is the mean molecular weight, $m_{\mathrm{H}}$ is the mass of a hydrogen atom, $v_w = 15$ km s$^{-1}$ is the assumed constant wind velocity, $R_{*}$ is the radius of the donor star where the stellar wind is launched, $r$ is the distance from the centre of the donor star, and $n(r)$ is the number density at distance $r$. The maximum number density at $r = R_{*}$ is set to $10^{15}$ cm$^{-3}$ due to the limitations of \C\ (see section 3.6 in part 2 of \C\ documentation `Hazy') . 

\begin{figure}
\centering
\includegraphics[width=0.47\textwidth]{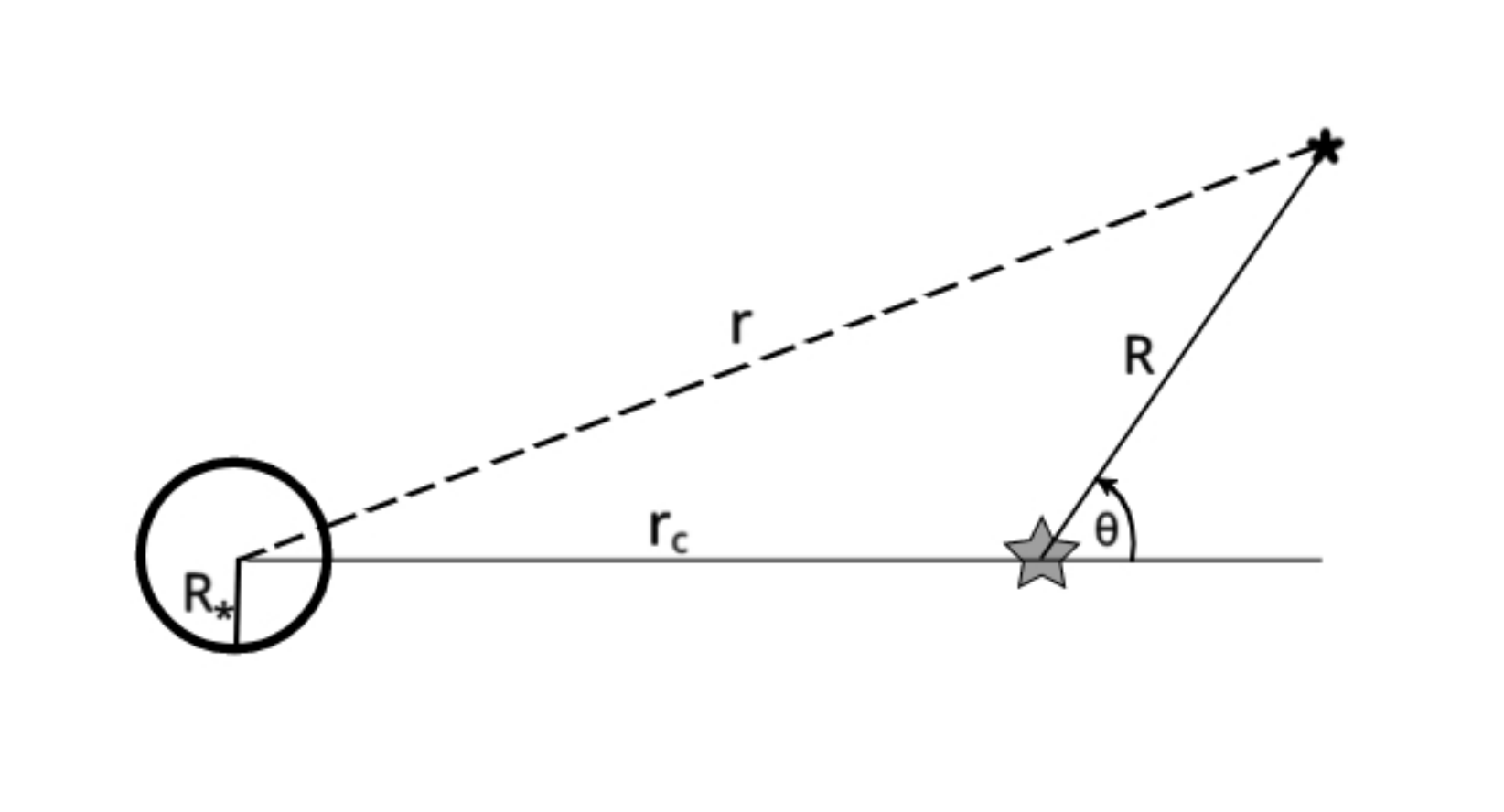}
\caption{The geometrical configuration of our simulations from \citet{Kuuttila21}. The giant star and the centre of the spherically symmetric density distribution is marked with the black open circle. The white dwarf is marked with the grey star at the distance $r_c$ from the density centre. A test particle at distance $r$ from the centre of the density distribution will have coordinates $(R, \theta)$ in the WD-centred reference system.}
\label{fig:system}
\end{figure}

Introducing an angle $\theta$ as the angle between the orbital plane and the line of sight from the WD (see Fig.~\ref{fig:system}), we can write the density distribution from the white dwarfs point of view as 
\begin{equation}\label{eq:nRT}
    n(R, \theta) =  \frac{n_c \, r_c^2}{R^2 + r_c^2 + 2 R r_c \mathrm{cos}\theta }  \left( 1 - \frac{R_*}{R^2 + r_c^2 + 2 R r_c \mathrm{cos}\theta} \right)^{-1}.
\end{equation}{}

In our simulations we assumed a solar metallicity from \citet{Grevesse10}, 
a diffuse background radiation field was included following \citep{Ostriker83, Ikeuchi86, Vedel94} with the cosmic microwave background included, and the cosmic rays were included in the calculations according to the mean ionisation rate of \citet{Indriolo07}. We ignored the radiation from the donor star, because the temperature of the donor star (3000 K) is low compared to the WD's temperature ($\gtrsim 10^5$ K) and thus does not significantly contribute to the ionization of the CSM.

\subsection{Binary parameters}

Our aim is to explore the  parameter space of symbiotic binaries and to study the dependence of the emission line content and luminosity  of symbiotic binaries on mass-loss rate of the donor star and the binary separation. To  this end, we need to fix several other parameters of the system.

We carried out our simulations for two different WD masses: 0.6 \msun\ and 1.0 \msun. The mass of the donor star was fixed to 1.0 \msun\ and 1.66 \msun, respectively, so that the mass ratio $q = M_{\mathrm{WD}}/M_{\mathrm{donor}}$ was fixed to 0.6. This was done because the wind Roche lobe overflow formalism of \citet{Abate13} used in this paper is based on hydrodynamical simulations \citep{Mohamed12}, where $q = 0.6$ was fixed. \citet{Abate13} introduced a $\propto q^2$ scaling to the accretion efficiency formula (see Eq.~(\ref{eq:beta_WRLOF})) based on the scaling in the Bondi-Hoyle-Lyttleton formalism, but as the validity of this scaling has not been verified yet, we focus only on the original $q = 0.6$. 
For both values of the donor star mass we assumed its temperature to be 3000 K and the radius to be 200 \rsun. The dust in the donor star wind was assumed to be carbon rich (see Eq.~(\ref{eq:dustradius})). 

We varied the mass-loss rate from the donor star from 10$^{-8}$ to 10$^{-5}$ \msun, which corresponds to the typical mass-loss rates from AGB stars \citep{Hoefner18}. 
The orbital separation between the donor star and the WD was varied from 2 to 14 au (period from 2.2 to 37 years). One should note that for separations $\lesssim 2$ AU the Roche lobe radius of the donor approaches the radius of the star and the accretion transitions to the traditional Roche lobe overflow. For example. for 2 AU separation the Roche lobe size of the donor is 182 \rsun, which is already smaller than the assumed donor size (200 \rsun). As our simulations focus on symbiotic binaries which operate in the wind accretion regime we do not consider the Roche lobe overflowing systems in this paper.

\subsection{Accretion rate}

Given the mass-loss rate from the donor star, we calculate the mass-accretion rate on the WD using the same method as in \citet{Abate13}. In this method the accretion rate is calculated with both the standard Bondi-Hoyle-Lyttleton (BHL) formalism \citep{Hoyle39,Bondi44} and the wind Roche lobe overflow (WRLOF) formalism \citep{Mohamed07, Mohamed12}, and the higher of the two is chosen. The maximum accretion rate is set to 50\% of the mass-loss rate. 

The accretion efficiency in the BHL formalism is given by \citep{Boffin88}:
\begin{equation}\label{eq:beta_BHL}
    \beta _{\mathrm{BHL}} = \frac{\alpha}{2 \, \sqrt{1-e^2}} \left( \frac{G \, M_{\mathrm{WD}}}{ r_{\mathrm{c}} \, v^2_{\mathrm{w}}} \right)^2 \left[ 1 + \left( \frac{v_{\mathrm{orb}}}{v_{\mathrm{w}}} \right)^2 \right]^{-3/2},
\end{equation}
where $\alpha = 1.5$ is a constant, $e$ is the eccentricity (assumed zero), and $v_{\mathrm{orb}}$ is the orbital velocity. 
The accretion efficiency in the WROLF formalism is given by \citep{Abate13}:
\begin{equation} \label{eq:beta_WRLOF}
    \beta _{\mathrm{WRLOF}} = \frac{25}{9} \, q^2 \, \left[ -0.284 \left( \frac{R_{\mathrm{d}}}{R_{\mathrm{L}}} \right)^2 + 0.918 \frac{R_{\mathrm{d}}}{R_{\mathrm{L}}} - 0.234 \right],
\end{equation}
where $q = M_{\mathrm{WD}}/M_{\mathrm{donor}}$ is the mass ratio, $R_{\mathrm{d}}$ is the dust formation radius, and $R_{\mathrm{L}}$ is the Roche lobe radius of the donor star \citep{Eggleton83}. 
The dust formation radius can be calculated by \citep{Hoefner07}:
\begin{equation} \label{eq:dustradius}
    R_{\mathrm{d}} = \frac{1}{2} R_{\mathrm{*}} \left( \frac{T_{\mathrm{d}}}{T_{\mathrm{eff}}} \right)^{-\frac{4+p}{2}},
\end{equation}
where $R_{\mathrm{*}}$ is the radius of the donor star, $T_{\mathrm{d}}$ is the dust condensation temperature, $T_{\mathrm{eff}}$ is the temperature of the donor star, and $p$ is a parameter characterising the opacity of the dust. For amorphous carbon dust grains $T_{\mathrm{d}} = 1500$ K and $p = 1$ \citep{Hoefner07}. 

In each of our simulations, for a given mass-loss rate and binary separation, we calculated the mass-accretion rate with the equations (\ref{eq:beta_BHL}) and (\ref{eq:beta_WRLOF}) and chose the higher value, i.e.
\begin{equation}
    \dot{M}_{\mathrm{acc}} = \mathrm{max} \left\{ \beta _{\mathrm{BHL}} (r_c), \beta _{\mathrm{WRLOF}} (r_c) \right\} \times \dot{M}_{\mathrm{loss}}
\end{equation}
 and then impose the condition that $\dot{M}_{\mathrm{acc}}<0.5\times \dot{M}_{\mathrm{loss}}$.

\subsection{Luminosity and temperature of the WD}

The temperature and luminosity of the WD are calculated  directly from the mass-accretion rate and the WD parameters. The calculations were done in two different ways, depending on whether the accretion rate was above the steady nuclear burning limit or below. 

If the accretion rate is high enough to sustain steady nuclear burning of hydrogen, 
we determined the colour temperature and the photospheric radius of the WD based on the calculations of \citet[][see Figs. 3 and 4]{Hachisu99b}, for a given WD mass and accretion rate. The spectrum of the WD was assumed to be black-body, so the luminosity was then calculated with Stefan-Boltzmann law. 

For lower accretion rates the total luminosity was assumed to be the accretion luminosity:
\begin{equation}
    L_{\mathrm{acc}} = \frac{G M_{\mathrm{WD}} \dot{M}}{R_{\mathrm{WD}}}.
\end{equation}
Half of this luminosity was assumed to be radiated by a geometrically thin and optically thick accretion disk \citep{Shakura73} with a multi-colour black-body spectrum \citep{Mitsuda84}. The maximum temperature of the disk is $T_{\mathrm{max}} = 0.488 \, T_{*}$, where 
\begin{equation}
    T_* = \left( \frac{3 G M_{\mathrm{WD}} \dot{M} }{8 \pi R_{\mathrm{WD}}^3 \sigma _{\mathrm{SB}}} \right)^{1/4},
\end{equation}
where $R_{\mathrm{WD}}$ is the radius of the WD and $\sigma _{\mathrm{SB}}$ is the Stefan-Boltzmann constant. 

The other half of the accretion luminosity was assumed to be radiated by a boundary layer (BL). For WDs with high accretion rates \citep[$\gtrsim$ 10$^{16}$ g s$^{-1}$;][]{Pringle79, Popham95}, the BL is optically thick with temperatures of $10^{5-6}$ K, while for lower accretion rates the BL is optically thin with temperatures up to $10^{8}$ K \citep{Pringle77, Popham95, Suleimanov14}. 
We have focused only on the high accretion rate regime, so we have assumed a BL emitting a black body spectrum with temperature of \citep{Frank02}
\begin{equation}
    T_{\mathrm{BL}} = T_{*} \left( \frac{3 G M_{\mathrm{WD}} \mu m_{\mathrm{H}} }{8 k R_{\mathrm{WD}} T_{*}} \right)^{1/8},
\end{equation}
where $\mu$ is the mean molecular weight and $k$ is the Boltzmann constant. 
This simple formula is in agreement with the theoretical BL simulations of \citet{Hertfelder13, Suleimanov14}. 

\begin{figure}
\centering
\includegraphics[width=0.47\textwidth]{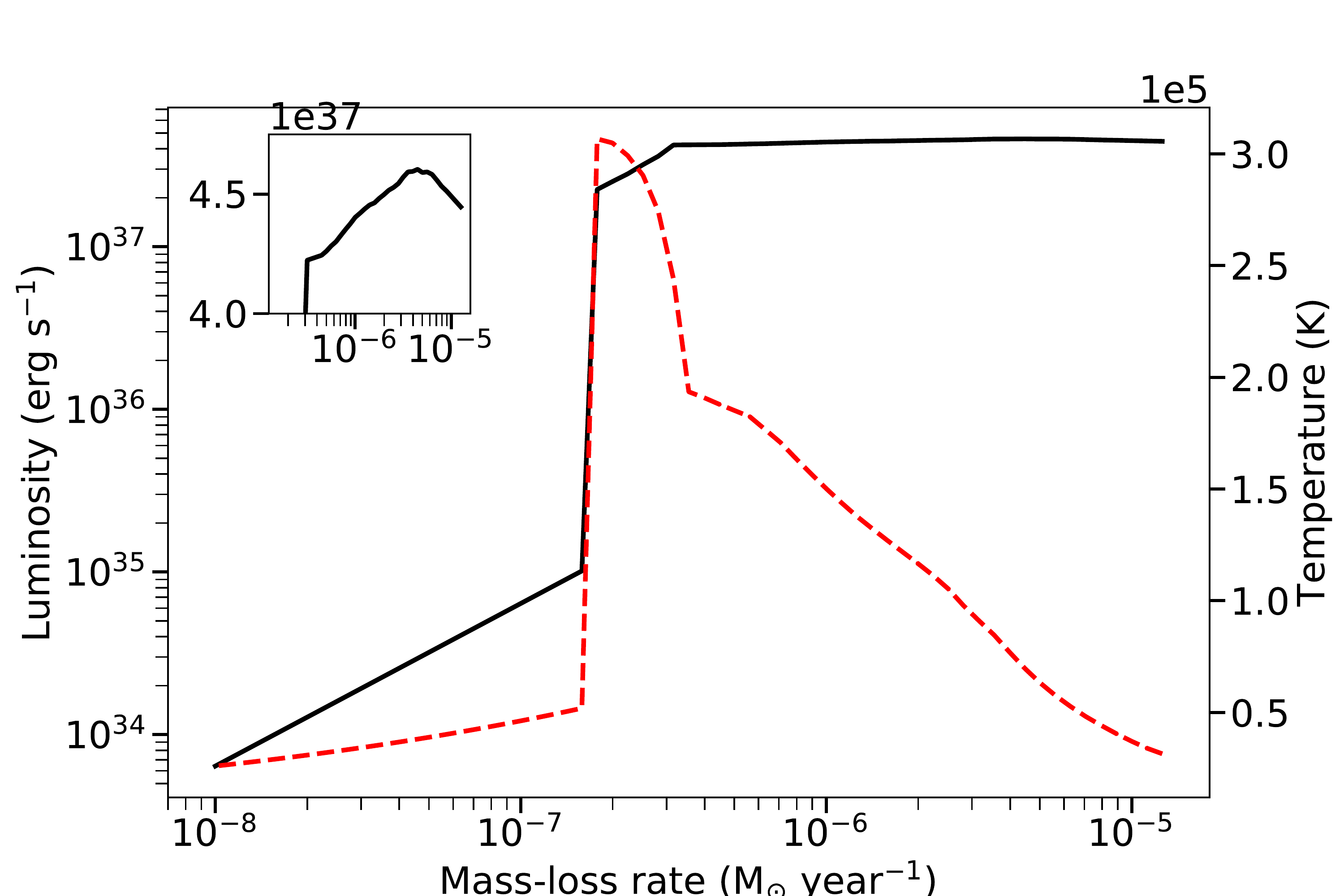}
\includegraphics[width=0.47\textwidth]{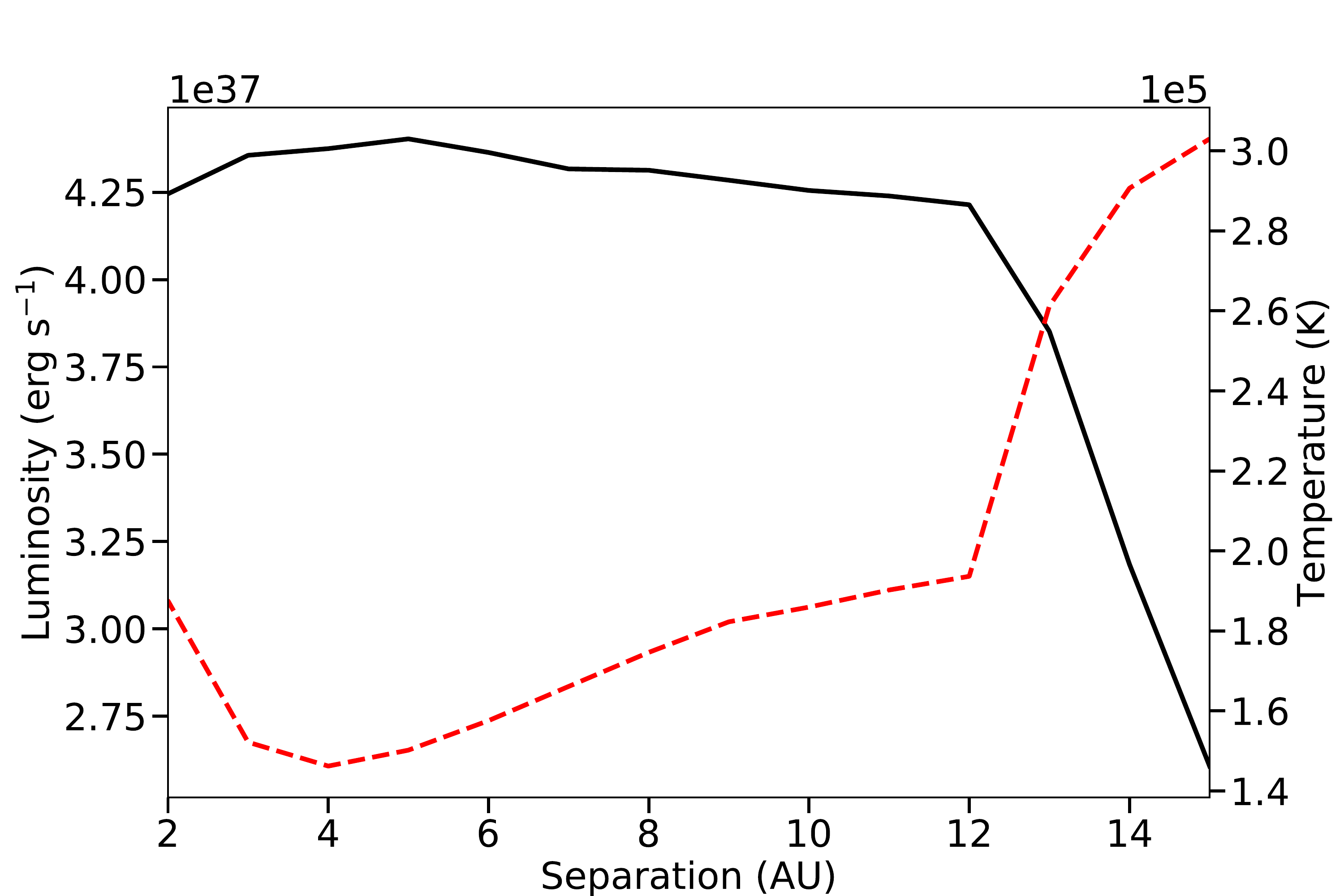}
\caption{ \textit{Upper panel:} The luminosity and temperature of the WD as a function of the mass-loss rate from the donor star with constant separation of 5 AU. The y-axis on the left shows the luminosity in erg s$^{-1}$ (black line) in logarithmic scale, and the y-axis on the right shows the temperature in K (red dashed line) in linear scale. The insert on the top left shows the luminosity in linear scale in the high mass-loss regime. The sharp rise in both temperature and luminosity at $\sim 2 \times 10^{-7}$ \msun\ yr$^{-1}$ is because of the onset of the nuclear burning on the WD. \newline
\textit{Lower panel:} The luminosity and temperature of the WD as a function of separation assuming a constant mass-loss rate of $10^{-6}$ \msun\ yr$^{-1}$. The y-axis on the left shows the luminosity (black line) in linear scale, and the y-axis on the right shows the temperature (red dashed line) in linear scale. In both panels, for low mass-loss rates with no steady nuclear burning the temperature shown is the maximum temperature of the accretion disk.}
\label{fig:temp_lum}
\end{figure}

In Fig.~\ref{fig:temp_lum} we show the temperature and the luminosity of the WD as a function of the mass-loss rate from the donor star (upper panel) and the orbital separation (lower panel). The mass of the WD in this figure is set to 0.6 \msun. 

We note that we did not take into consideration the impact of the classical nova explosions on the structure of the accretion disk and CSM. In the major part of the parameter space classical nova explosions are of importance only during a small fraction of time.

\section{Results}\label{sec:results}

Using our model described in Sec.~\ref{sec:model}, we can calculate self-consistently the two dimensional temperature, ionization and emission structure of the CSM around symbiotic binaries. With this method we can predict the line emission spectrum and how it depends on the mass-loss rate from the donor star and the orbital separation of the binary system. 

Here we describe  results for a 0.6 \msun\ WD, results for a 1.0 \msun\ WD are presented  the Appendix. The results for both values of the WD masses are qualitatively similar.  

\subsection{Absorption of the WD emission by CSM}

The emission from symbiotic binaries is often thought to be heavily obscured by the dense neutral wind from the donor star. Indeed, the X-ray emission observed from symbiotic systems is often  heavily absorbed, but this is mostly due to the low colour temperature of the WD and the ISM absorption \citep{Nielsen15}. With our simulations we can show that the CSM in symbiotic binaries is in fact mostly ionized and does not obscure the WD emission \citep[see also][]{Kuuttila21}. 

\begin{figure}
\centering
\includegraphics[width=0.47\textwidth]{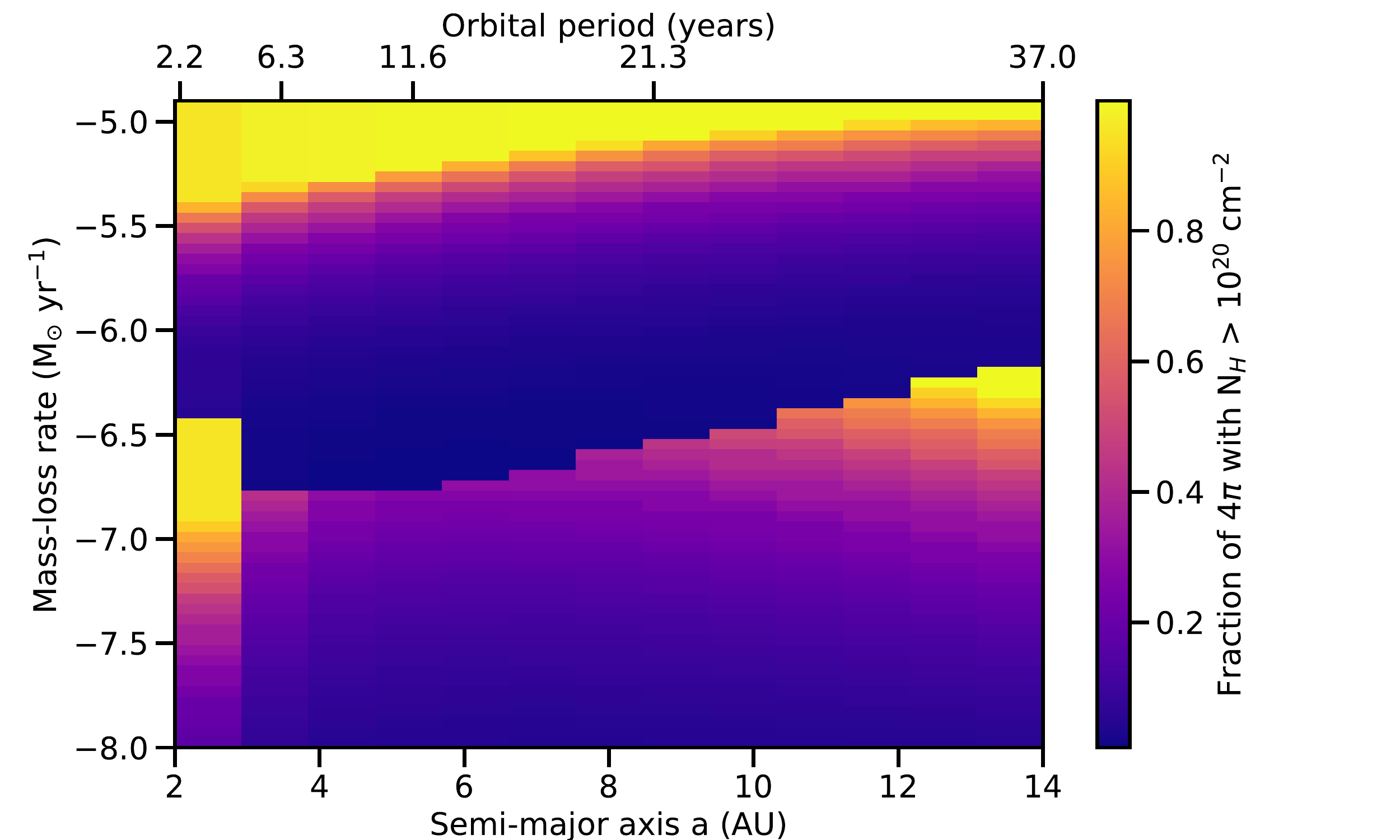}
\caption{The fraction of the solid angle around the WD filled with the neutral CSM. The x-axis shows the orbital separation at the bottom and the Keplerian orbital period on top, and the y-axis shows the mass-loss rate from the donor star. The colour in this figure illustrates the fraction of the solid angle, where N$_{H} > 10^{20}$ cm$^{-2}$, minus the fraction of the sky covered by the donor star. The jump between 10$^{-7}$ and 10$^{-6.5}$ \msun\ yr$^{-1}$ is caused by the onset of steady nuclear burning on the WD surface. }
\label{fig:columnDepth}
\end{figure}

For each simulation, we have calculated the neutral hydrogen column density N$_{H}$ along each line of sight from the WD, i.e. as a function of $\theta$ (see Fig.~\ref{fig:system}). These results are summarised in Fig.~\ref{fig:columnDepth}, where we show the fraction of the full solid angle 4$\pi$ around the WD with the column density of neutral hydrogen N$_{\mathrm{H}} > 10^{20}$ cm$^{-2}$, minus the solid angle occupied by the donor star, i.e.:
\begin{equation}
    f = \frac{\Omega _{\mathrm{N_H} > 10^{20}} - \Omega _{\mathrm{donor}}}{4 \pi},
\end{equation}
as a function of the orbital separation and mass-loss rate from the donor. 
This figure illustrates the fraction of the sky, from the WD's point of view, where the wind is mostly ionized and thus transparent. For the observer this plot illustrates the probability of observing the WD with a large amount of neutral absorption due to the CSM.

From this figure one can see that the gas around the WD is mostly ionized in large fraction of the parameter space. As it could have been expected, the CSM becomes mostly neutral and obscuring for very high mass-loss rates $\sim 10^{-5}$ \msun\ yr$^{-1}$. The CSM absorption is also significant at the intermediate mass accretion rates and large orbital separations, when the steady nuclear burning has not started and the accretion luminosity is insufficient in ionizing the dense wind at large distances from the WD. For small orbital separation near 2 AU and below the nuclear burning limit, the wind is also mostly neutral due to the WD's proximity to the donor star.

\begin{figure}
\centering
\includegraphics[width=0.47\textwidth]{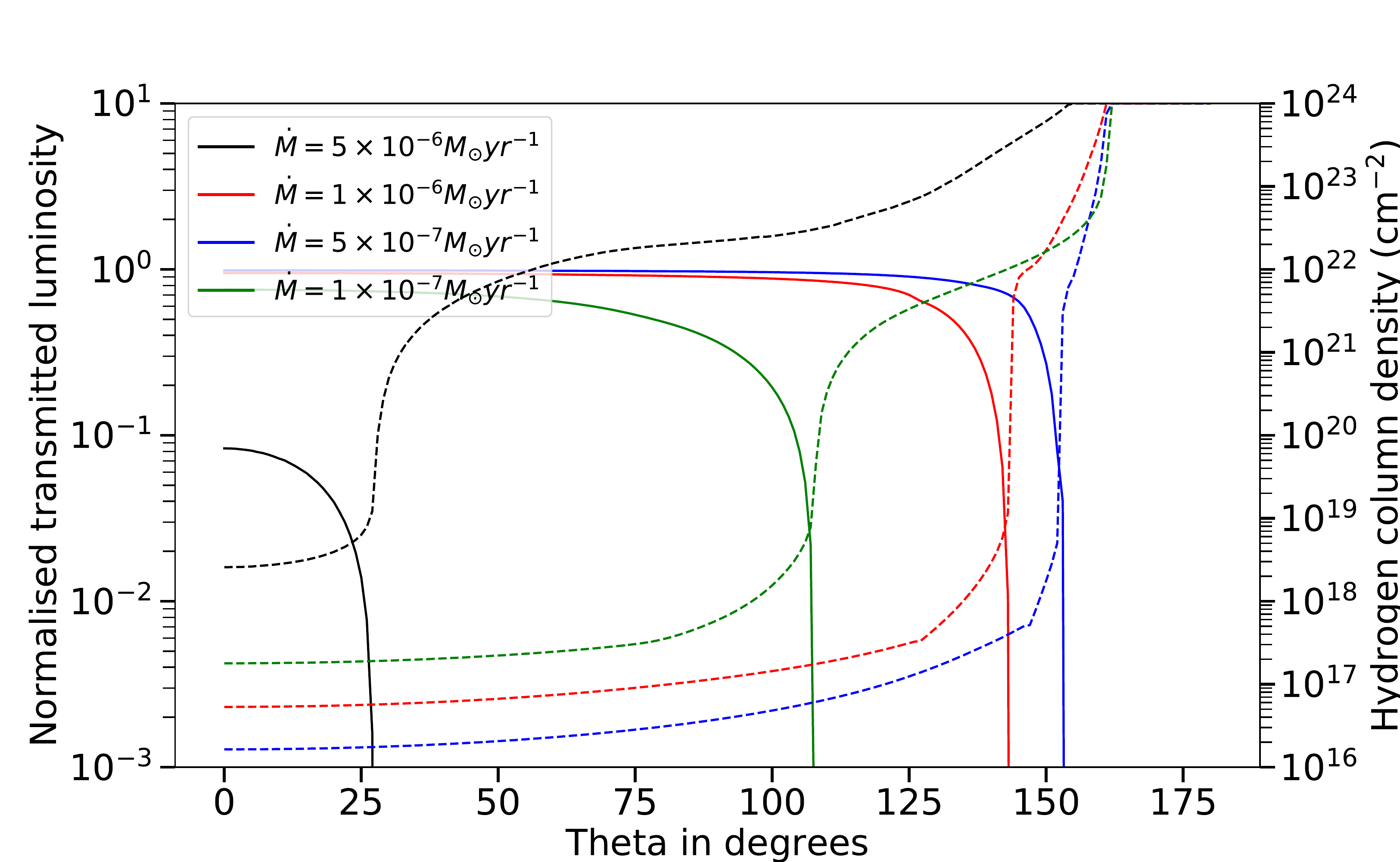}
\caption{The transmitted luminosity of the WD (solid lines) and the CSM neutral hydrogen column density (dashed lines) as a function of the angle $\theta$. The y-axis on the left side show the transmitted luminosity normalised to the initial WD luminosity, and the y-axis on the right show the column density in units of cm$^{-2}$. The black, red, blue, and green lines show the results for separation of 3 AU and mass-loss rates of $5 \times 10^{-6}$, $1 \times 10^{-6}$, $5 \times 10^{-7}$, and $1 \times 10^{-7}$ \msun\ yr$^{-1}$, respectively. }
\label{fig:attenuation_angle}
\end{figure}

Knowing the distribution of ionization fraction around the WD, we can also study the attenuation of the WD emission by the circumstellar medium as a function of the angle $\theta$. In Fig.~\ref{fig:attenuation_angle} we show the normalised, transmitted bolometric luminosity of the WD for four different mass-loss values. 
Here one can see that in the steady burning regime, i.e. mass-loss rates of $1 \times 10^{-6}$ and $5 \times 10^{-7}$ \msun\ yr$^{-1}$, the CSM is mostly ionised and the emission from the WD is attenuated only near the donor star. For higher mass-loss rates, the low colour temperature of the WD and the high density mean that the WD emission is highly attenuated, while for mass-loss rates below the steady burning limit the WD emission is sufficient to ionize the wind in the significant fraction of the volume.

\begin{figure}
\centering
\includegraphics[width=0.47\textwidth]{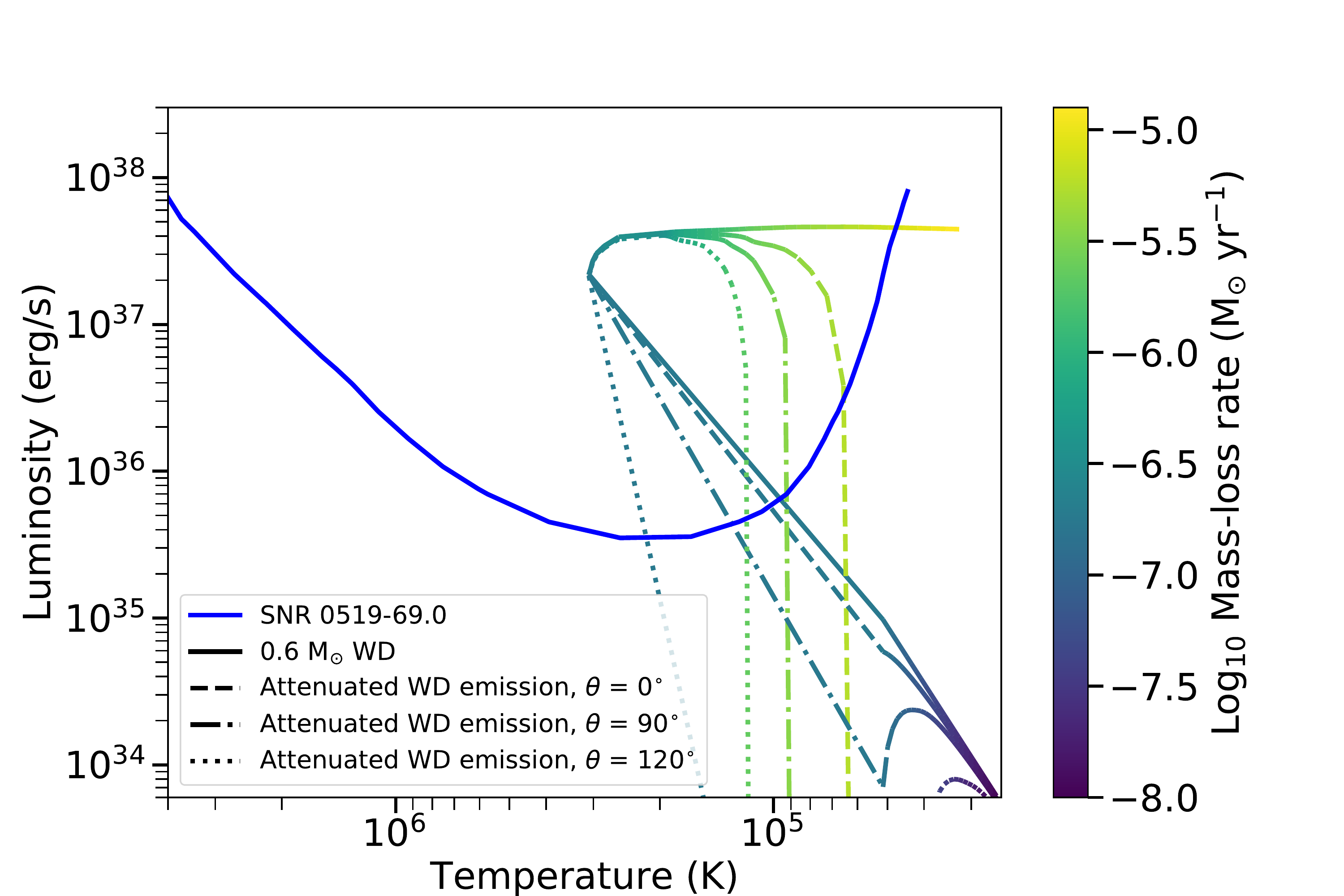}
\caption{The temperature and attenuated luminosity of the WD as a function of mass-loss rate. The solid coloured line shows the initial WD temperature and luminosity, while the dashed, dash-dotted, and dotted lines show the attenuated emission for angles 0$^{\circ}$, 90$^{\circ}$, and 120$^{\circ}$, respectively. The mass of the WD is 0.6 \msun\ and the orbital separation is 3 AU.
The solid blue line shows the upper limit on the temperature and luminosity of the progenitor of type Ia SNR 0519-69.0 from \citet{Kuuttila19}. }
\label{fig:lum_temp_attenuation}
\end{figure}

We can also study the attenuation of the WD emission in the context of type Ia supernova remnants. Accreting and nuclear burning WDs are expected to create extended Str\"omgren spheres around them, which can be used to constrain their contribution to the type Ia progenitors \citep[e.g.][]{Kuuttila19}. Should the WD emission be obscured by the CSM around it, the resulting brightness of the ionized nebula and thus the progenitor constraints may be weaker. 
In Fig.~\ref{fig:lum_temp_attenuation} we compare the attenuation of the WD emission to the upper limits on the progenitor of SNR 0519-69.0 from \citet{Kuuttila19}. In this figure we show the initial WD temperature and luminosity (see Sec.~\ref{sec:model}) and the attenuated emission for three different lines of sight ($\theta = 0^{\circ}, 90^{\circ}, 120^{\circ}$) as a function of the mass-loss rate. 
From this figure one can see that for a 0.6 \msun WD the temperatures and luminosities below the steady burning limit are below the upper limit derived for SNR 0519-69.0. 
In the steady burning regime the temperature and luminosities are above the upper limit for SNR 0519-69.0 for all angles. For very high mass-loss rates ($\gtrsim$ 10$^{-5.5}$ \msun yr$^{-1}$) the WD emission becomes obscured for all angles, with the transition happening at slightly lower mass-loss rates for higher angles.

\subsection{Line emission}

Emission lines are the main diagnostic tools for  symbiotic binaries -- they  are often bright in H Balmer lines, He \textsc{ii} 4686\AA, and various other emission lines. Notably, high ionization state iron lines, e.g. [Fe \textsc{x}] 6374\AA, have also been often  observed.
Below we model the emission line spectra of these objects and  investigate their dependence on the mass-loss rate of the donor star and the binary orbital separation. 

\begin{figure*}
\centering
\includegraphics[width=\textwidth]{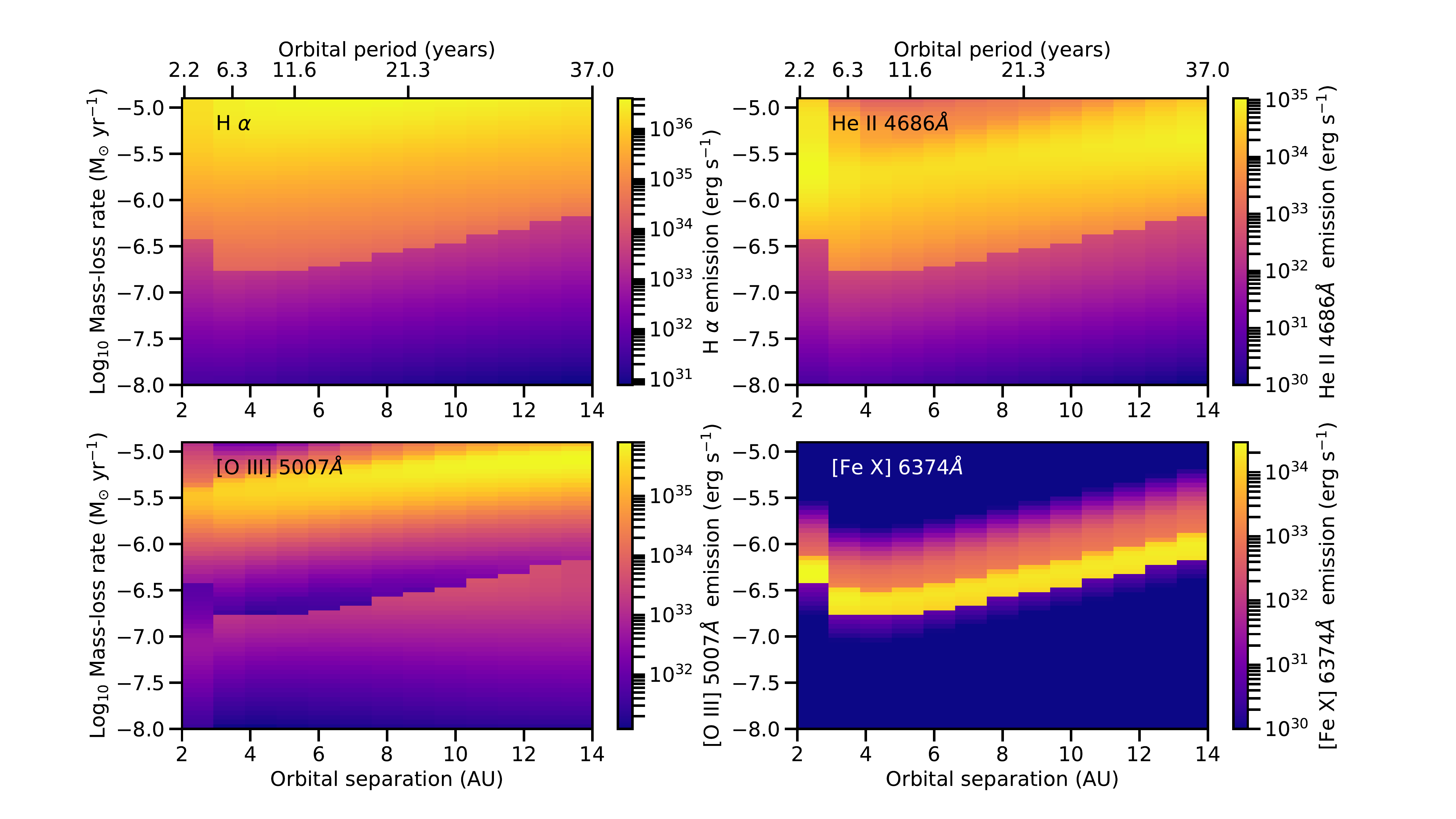}
\caption{The line luminosity (in erg s$^{-1}$) for four different emission lines as predicted by the \C\ simulations. The x-axis shows the orbital separation $r_c$ in AUs and the Keplerian orbital period in years for a WD mass 0.6 \msun. The colour shows the line luminosity for H$\alpha$ (top left), He \textsc{ii} 4686 (top right), [O \textsc{iii}] 5007 (bottom left), and [Fe \textsc{x}] 6374 \AA\ (bottom right). }
\label{fig:grid}
\end{figure*}

From our simulations we obtain the volume emissivity of each line, which is then integrated over the simulated volume to get the total line luminosity, as also explained in \citet{Kuuttila21}. In Fig.~\ref{fig:grid} we show the line luminosity for four different emission lines: H$\alpha$, He \textsc{ii} 4686, [O \textsc{iii}] 5007, and [Fe \textsc{x}] 6374 \AA, as a function of the mass-loss rate and the orbital separation. The onset of nuclear burning can be seen as jump in the line luminosities between 10$^{-7}$ and 10$^{-6.5}$ \msun\ yr$^{-1}$. 

From Fig.~\ref{fig:grid} one can see that the H$\alpha$ emission increases monotonically with the mass-loss rate with little dependence on the orbital separation. The situation is almost the same for He \textsc{ii} 4686\AA, except for the decreasing emission at very high mass-loss rates, when the photosphere of the WD is inflated and the colour temperature decreasing. 

For [O \textsc{iii}] 5007\AA\ the emission first increases with the increasing mass-loss rate, but then decreases sharply at the onset of nuclear burning, when the CSM becomes too highly ionized (see Sec.~\ref{sec:forbiddenlines}). Then the emission increases again with increasing mass-loss rates until the density becomes very high. 
For [Fe \textsc{x}] 6374 \AA\ the emission is sharply peaked at the peak of the steady nuclear burning. Without nuclear burning the WD is not hot and luminous enough to produce appreciable amounts of [Fe \textsc{x}] emission, and with high mass-loss rates ($\gtrsim 10^{-6}$ \msun yr$^{-1}$) the decreasing colour temperature of the WD is not enough to produce highly ionized iron. Thus [Fe \textsc{x}] 6374 \AA\  emission line is a clear signature of an accreting and steadily burning WD within the stability strip.

\begin{figure*}
\centering
\includegraphics[width=\textwidth]{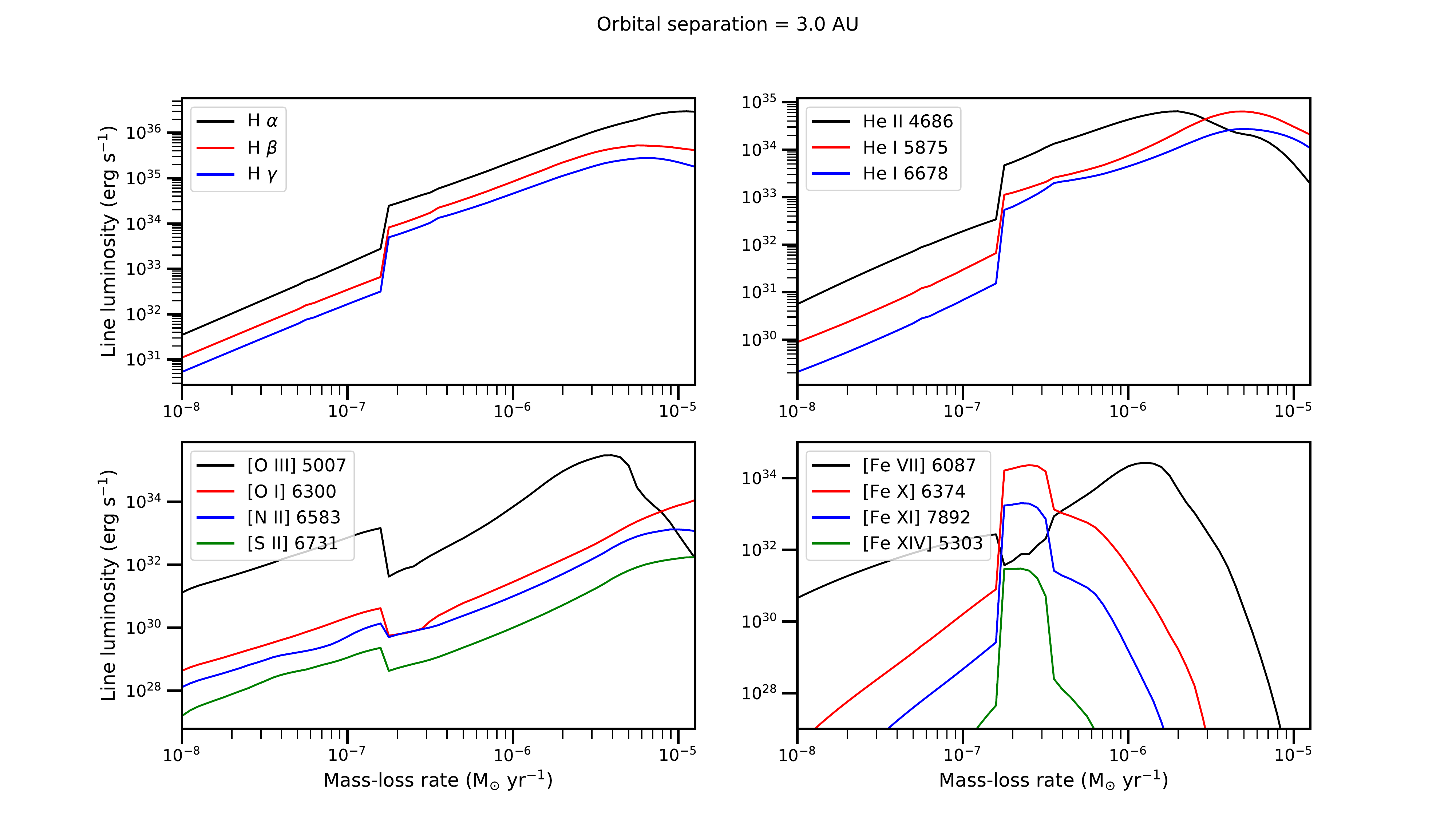}
\caption{The line luminosity (in erg s$^{-1}$) as a function of the mass-loss rate for various emission lines. The orbital separation is set to 3 AU. The top left panel shows the line luminosity for H$\alpha$ (black line), H$\beta$ (red line), and H$\gamma$ (blue line). The top right shows the He \textsc{ii} 4686 (black), He \textsc{i} 5875 (red), and He \textsc{i} 6678 \AA\ lines. The bottom left shows the common forbidden [O \textsc{iii}] 5007 (black), [O \textsc{i}] 6300 (red), [N \textsc{ii}] 6583 (blue), and [S \textsc{ii}] 6731 \AA\ (green) emission lines. The bottom right panel shows various iron lines: [Fe \textsc{vii}] 6087 (black), [Fe \textsc{x}] 6374 (red), [Fe \textsc{xi}] 7892 (blue), and [Fe \textsc{xiv}] 5303 \AA\ (green).}
\label{fig:lineEmission_3.0}
\end{figure*}

\begin{figure*}
\centering
\includegraphics[width=\textwidth]{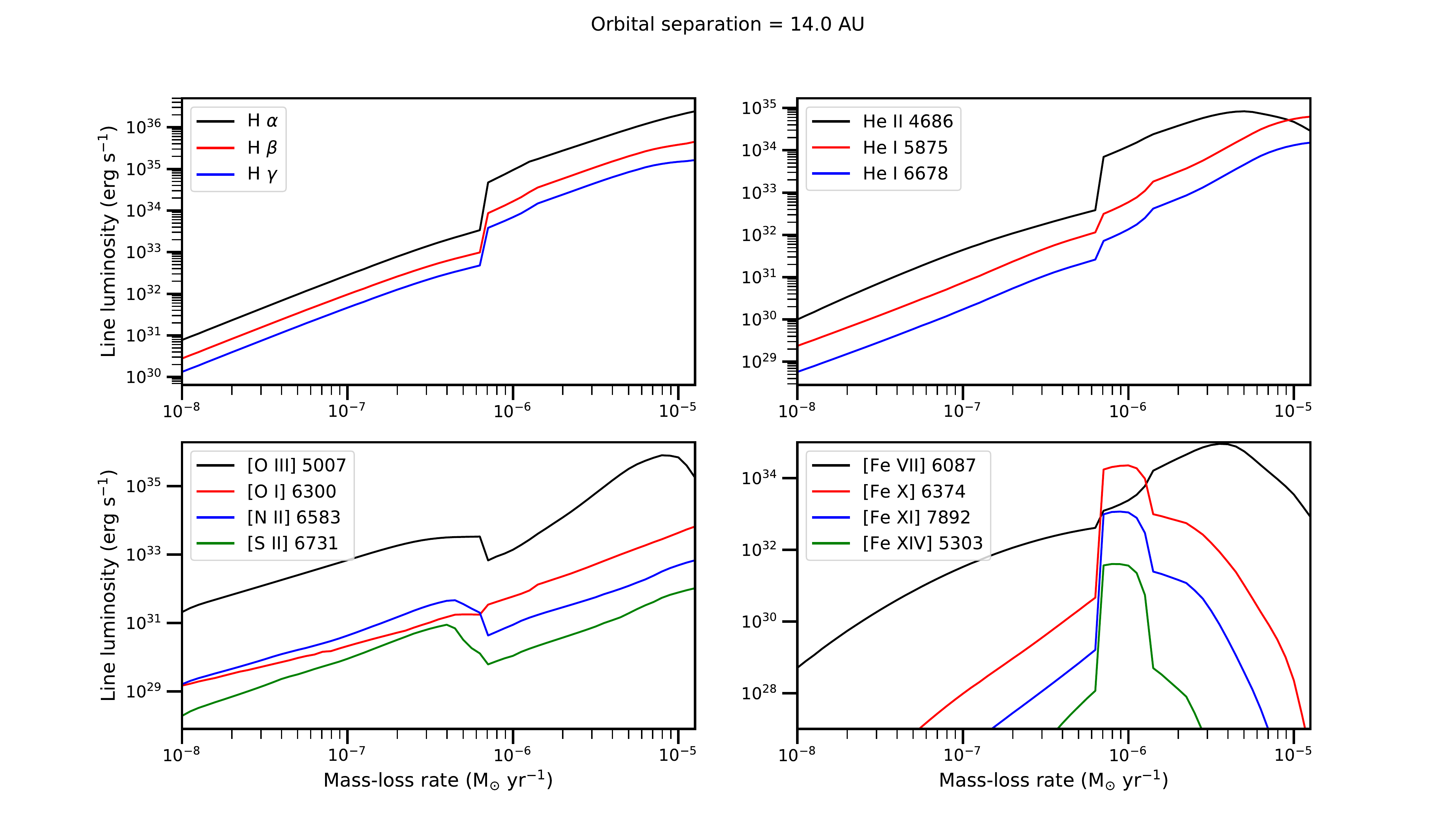}
\caption{Same as Fig.~\ref{fig:lineEmission_3.0} but for separation of 14 AU. }
\label{fig:lineEmission_14.0}
\end{figure*}

In Fig.~\ref{fig:lineEmission_3.0} we show the luminosity of a number of   lines as a function of mass-loss rate for orbital separation of 3 AU, and in Fig.~\ref{fig:lineEmission_14.0} we show the same lines for the orbital separation of 14 AU. The lines included in these figures are: H$\alpha$, H$\beta$, and H$\gamma$ (top left), He \textsc{i} 5875, 6678 and He \textsc{ii} 4686 \AA\ (top right), [O \textsc{iii}] 5007, [O \textsc{i}] 6300, [N \textsc{ii}] 6583, and [S \textsc{ii}] 6731 \AA\ (bottom left), and [Fe \textsc{vii}] 6087, [Fe \textsc{x}] 6374, [Fe \textsc{xi}] 7892, and [Fe \textsc{xiv}] 5303 \AA\ (bottom right). 

The luminosity of emission lines from high ionisation state of iron  is strongly peaked and the shape of these curves follow closely the temperature dependence (see Fig.~\ref{fig:temp_lum}). For [Fe \textsc{vii}] emission there is a decrease at the same mass-loss rates where higher ionization state lines peak, because the gas is more strongly ionized. 

Overall, from Figs.~\ref{fig:grid}, \ref{fig:lineEmission_3.0}, and \ref{fig:lineEmission_14.0} one can see that there is no significant qualitative dependence on the orbital separation. The pattern of the mass-loss dependence of the line luminosity  is qualitatively same for different separations, but with the curves for larger separation shifted towards higher mass-loss rates, because of the smaller mass-accretion efficiency. 

\subsubsection{Forbidden emission lines}\label{sec:forbiddenlines}

In some symbiotic binaries none of the common nebular forbidden emission lines are observed \citep[e.g. LIN358;][]{Kuuttila21} which is commonly ascribed to the high density of the gas surrounding the WD. But as shown in Fig.~\ref{fig:lineEmission_3.0}, there is still some non-negligible luminosity  of e.g. [O \textsc{iii}] 5007 emission even at high mass-loss rates, although the [O \textsc{iii}] 5007 / H$\beta$ ratio is small. 

Interestingly, the luminosity of some of the  forbidden lines shown in Fig.~\ref{fig:lineEmission_3.0} increase  with increasing mass-loss rate, except for very high mass-loss rates. The sharp decrease in the line emission at about $2 \times 10^{-7}$ \msun yr$^{-1}$ is due to the increased ionization state of the gas at the onset of the nuclear burning, not due to the increased density. 
Thus, we conclude that in some cases the lack of detected forbidden  emission lines may be related also to the high ionization state and not only to the high density of the gas.  For some of the lines, however,  the gas density exceeds  the critical density, for example for [O \textsc{iii}] 5007 \AA, which ratio to H $\beta$ is very small. Overall, due to combination of ionization stae and gas density effects,  the forbidden line emission is quite faint in these systems.

\subsubsection{UV Lines}

\begin{figure}
\centering
\includegraphics[width=0.47\textwidth]{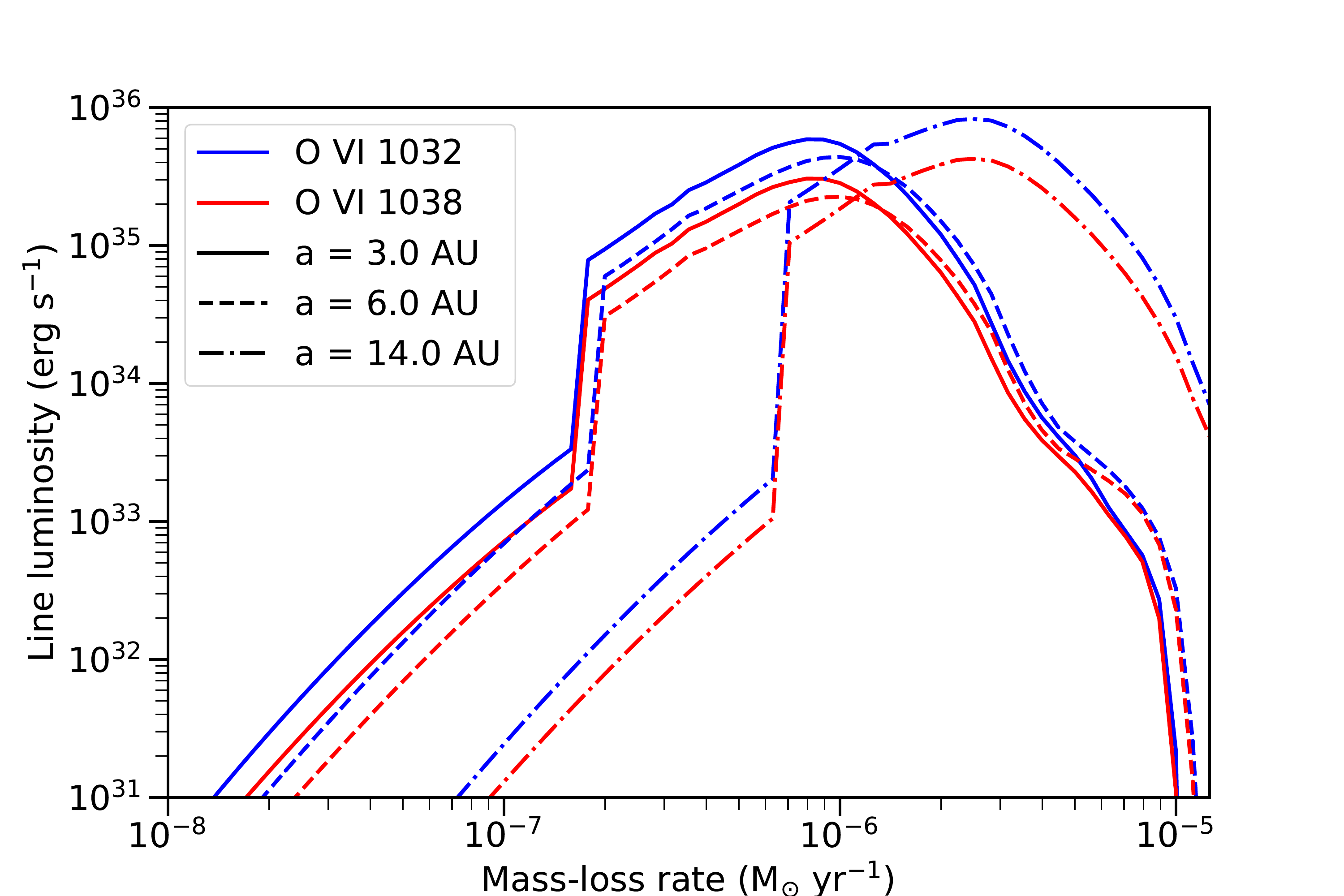}
\caption{The line luminosities of O \textsc{vi} 1032 \AA\ (blue), 1038 \AA\ (red) emission lines as a function of mass-loss rate for orbital separations of 3.0 (solid lines), 6.0 (dashed lines), and 14.0 AU (dash-dotted lines). }
\label{fig:UVlines}
\end{figure}

In Fig.~\ref{fig:UVlines} we show the line luminosities of the O \textsc{vi} $\lambda \lambda$1032, 1038 doublet lines. These lines are of special importance in symbiotic binaries due to the Raman scattering, which produces the broad emission bands at 6830 \AA\ and 7082 \AA. In this process the the O \textsc{vi} photons are absorbed by a ground state hydrogen leading to an intermediate state, from where a photon is emitted leaving the hydrogen in an excited state \citep{Schmid89}. 
This process is observed almost exclusively in symbiotic binaries, because it requires a hot ionizing source capable of producing O \textsc{vi} in the vicinity of a large amount of neutral hydrogen, with column densities of $\gtrsim 10^{23}$ cm$^{-2}$. 

The Raman scattering process is dependent on the geometry of the binary system and cannot be studied with \C\ but the typical scattering efficiencies from simultaneous UV and optical observations are 5\% -- 15\% for the $\lambda 1032 \rightarrow \lambda 6825$ and 2\% -- 12\% for the $\lambda 1038 \rightarrow \lambda 7082$ processes \citep{Espey95, Schmid99, Birriel00}. 

As one can see from Fig.~\ref{fig:UVlines} the O \textsc{vi} $\lambda \lambda$1032, 1038 lines are heavily dependent on the mass-loss rate, and they peak in the steady nuclear burning regime where the temperature and luminosity of the WD are the highest. There is no strong dependence on the orbital separation: for 3 AU and 6 AU separations the lines are almost the same, with the latter yielding slightly lower luminosities. For separation of 14 AU the shape of the luminosity curve is about the same, but it is shifted towards higher mass-loss rates due to the decreasing mass-accretion efficiency.

\subsection{Line ratios}

Various emission line ratios have been used in the past to characterise emission line objects and separate symbiotic binaries from e.g. planetary nebulae, Be stars and young stellar objects \citep[e.g.][]{Gutierrez-Moreno95, Rodriguez-Flores14, Ilkiewicz17}. 
We have investigated how our emission models of symbiotic binaries fit to these previously presented line ratio classification diagrams and how the line ratios depend on the mass-loss rate and orbital separation.

\begin{figure}
\centering
\includegraphics[width=0.47\textwidth]{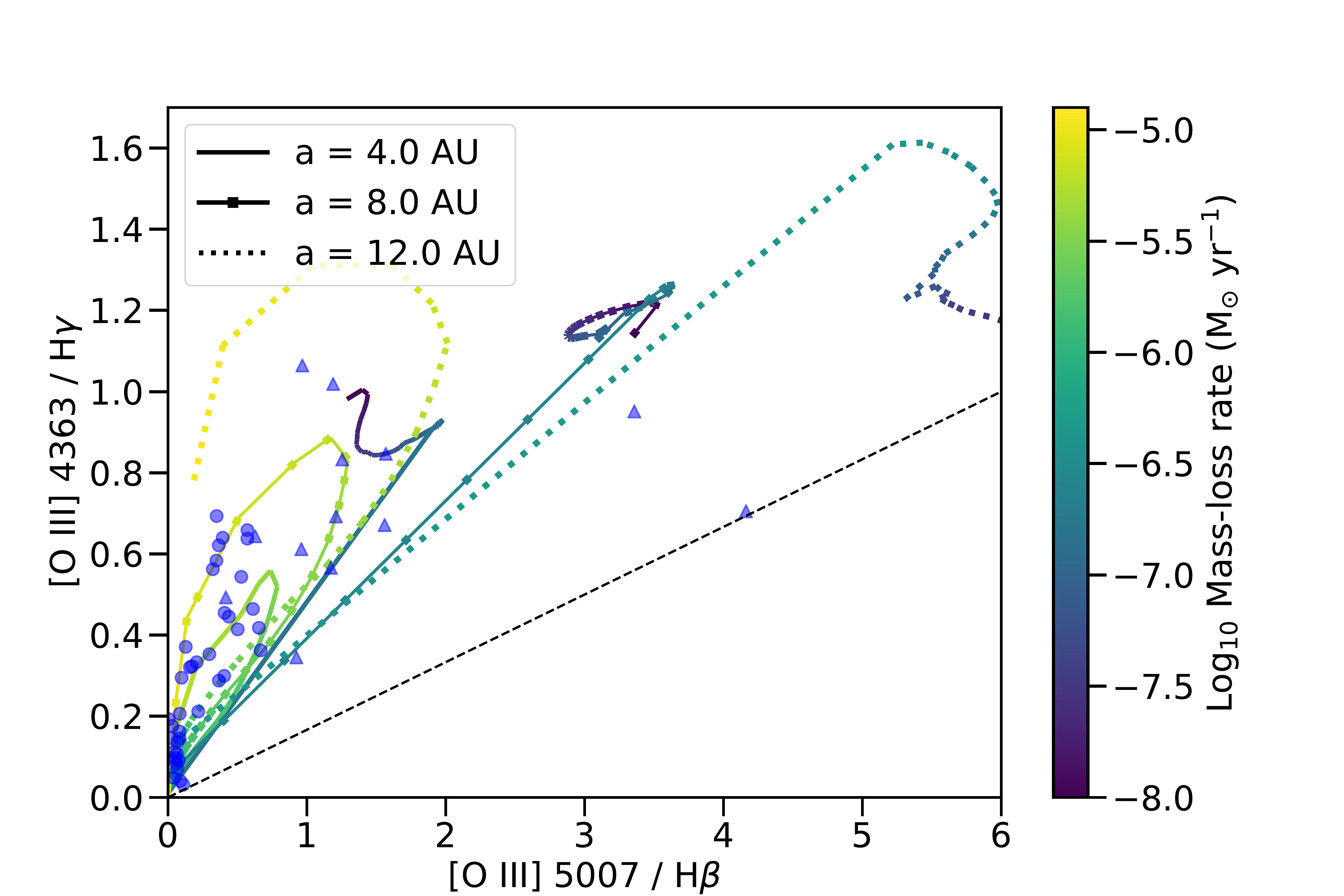}
\caption{[O \textsc{iii}] 5007/H$\beta$ vs. [O \textsc{iii}] 4363/H$\gamma$ line ratio diagram of \citet{Gutierrez-Moreno95}. The solid, solid-dotted, and the dotted lines show the line ratios for 4, 8, and 12 AU separations, respectively. The colour of the lines show the mass-loss rate according to the colourbar on the right. The blue circles show the S-type and the blue triangles show the D-type symbiotic binaries. The black dashed line shows the classification criterion of \citet{Gutierrez-Moreno95}: objects above this line are classified as symbiotic binaries. }
\label{fig:lineDiagram_1}
\end{figure}

In Figure~\ref{fig:lineDiagram_1} we show our models in the [O\textsc{iii}] diagnostic diagram of \citep{Gutierrez-Moreno95}. Shown in this figure is  the [O \textsc{iii}] 5007/H$\beta$ vs. [O \textsc{iii}] 4363/H$\gamma$ line ratios for three orbital separations: 4, 8, and 12 AU, while the colour of the line shows the mass-loss rate of the donor star. All three lines lines start with low mass-loss rates from the top right part of the figure, i.e. high [O \textsc{iii}] 5007/H$\beta$ and [O \textsc{iii}] 4363/H$\gamma$ ratios. With the start of the nuclear burning both of these ratios decrease close to zero and then with form a loop with increasing mass-loss rate. The behaviour with all separations is similar, but there is larger variation in the line ratios with larger separations. For comparison, we show the observed line ratios of various symbiotic binaries in this figure: the blue circles denote the S-type and the blue triangles mark the D-type binaries \citep{Mikolajewska97, Pereira98, Belczynski00, Luna05}. All the S-type symbiotics lie in the lower left corner of this figure, with smaller separations, while the D-type binaries have larger line ratios indicating larger separations, as expected \citep{Allen84, Gromadzki13}. Also shown in this figure with a black dashed line is the criteria of \citet{Gutierrez-Moreno95} used to separate symbiotics from planetary nebulae; all our models lie above this line where the symbiotic binaries are.

\begin{figure*}
\centering
\includegraphics[width=0.47\textwidth]{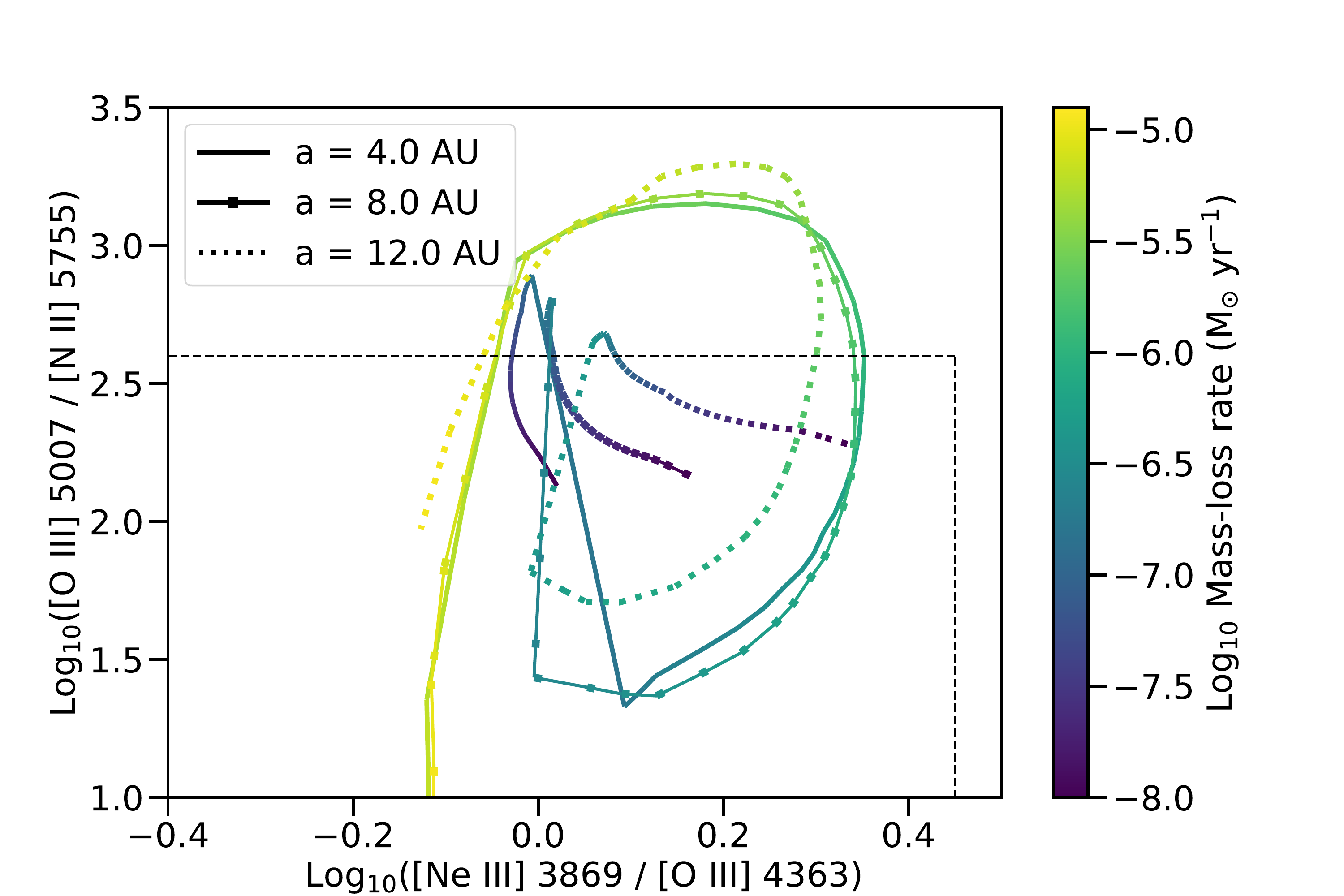}
\includegraphics[width=0.47\textwidth]{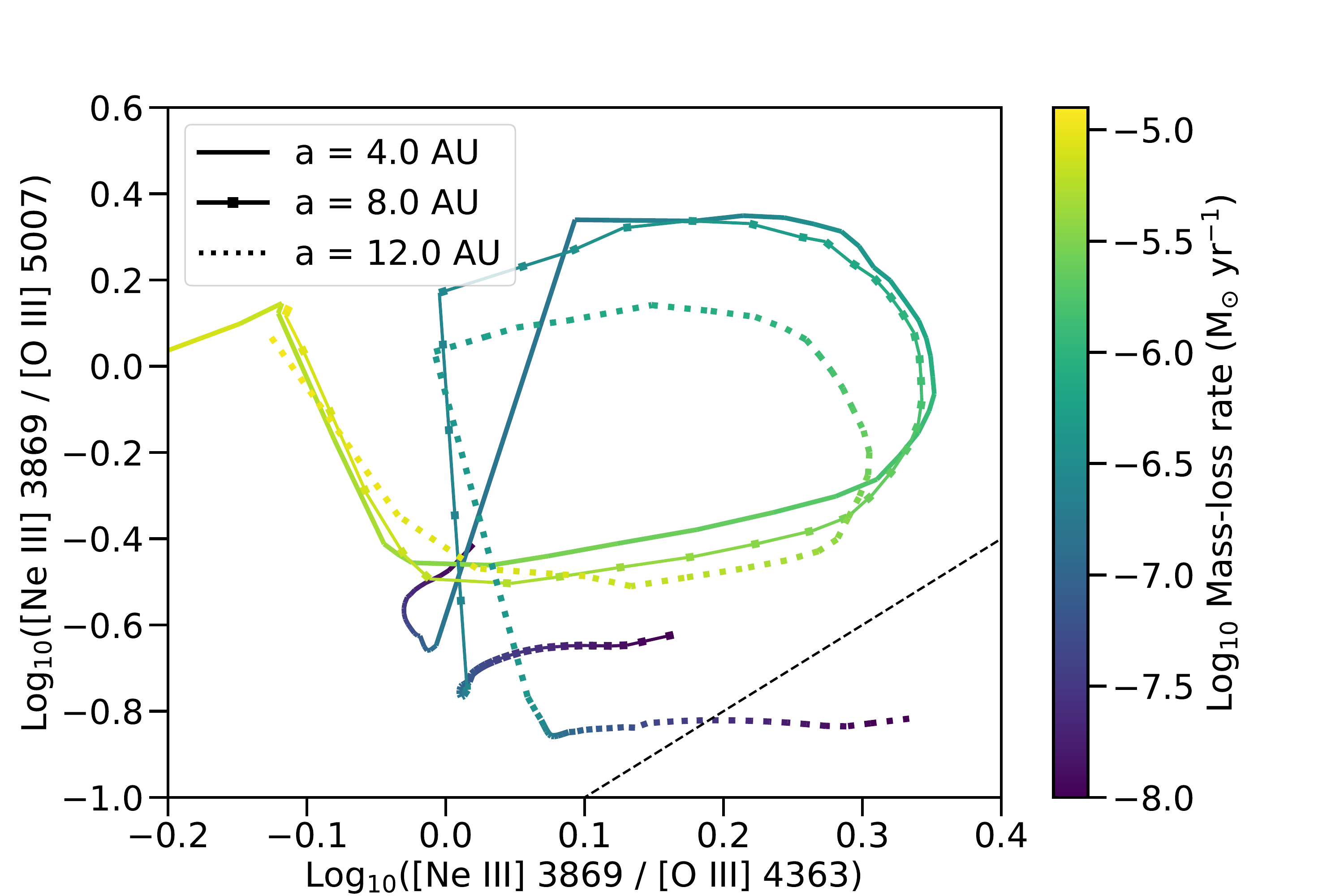}
\caption{Two different line ratio diagrams of \citet{Ilkiewicz17}. 
Shown on the left is the [Ne \textsc{iii}] 3869/[O \textsc{iii}] 4363 vs. [O \textsc{iii}] 5007/[N \textsc{ii}] 5755 diagram and shown on the right is the [Ne \textsc{iii}] 3869/[O \textsc{iii}] 4363 vs. [Ne \textsc{iii}] 3869/[O \textsc{iii}] 5007 diagram. 
The solid, solid-dotted, and the dotted lines show the line ratios for 4, 8, and 12 AU separations, respectively. The colour of the lines show the mass-loss rate according to the colourbar on the right. The black dashed lines show the the criteria for symbiotic binaries proposed by \citet{Ilkiewicz17}.}
\label{fig:lineDiagram_2}
\end{figure*}

\citet{Ilkiewicz17} studied the line ratios of many observed symbiotic binaries and planetary nebulae and proposed several new line ratio diagrams to distinguish between these two types of sources. We investigated two of the most promising diagrams in the context of our simulations. 
In Figure~\ref{fig:lineDiagram_2} we show the [Ne \textsc{iii}] 3869/[O \textsc{iii}] 4363 vs. [O \textsc{iii}] 5007/[N \textsc{ii}] 5755 and the [Ne \textsc{iii}] 3869/[O \textsc{iii}] 4363 vs. [Ne \textsc{iii}] 3869/[O \textsc{iii}] 5007 line ratio diagrams in a similar manner to Fig.~\ref{fig:lineDiagram_1}. Here one can see that there is no strong dependence on the orbital separation. Especially in higher mass-loss rates all three separations form a similar curve in this figure. Below the steady burning limit the ratios do not change much and lie close to each other making it difficult to infer the binary parameters from these line ratios. 
Also included in this figure is the classification criteria of \citet{Ilkiewicz17}. In the right panel of Fig.~\ref{fig:lineDiagram_2} the models lie in the region of symbiotics, except for the small part of the highest separation and lowest mass-loss rate model. In the left panel all the models lie partly outside the region of symbiotic binaries, especially for $\dot{M} > 10^{-6}$ \msun yr$^{-1}$.

\begin{figure}
\centering
\includegraphics[width=0.47\textwidth]{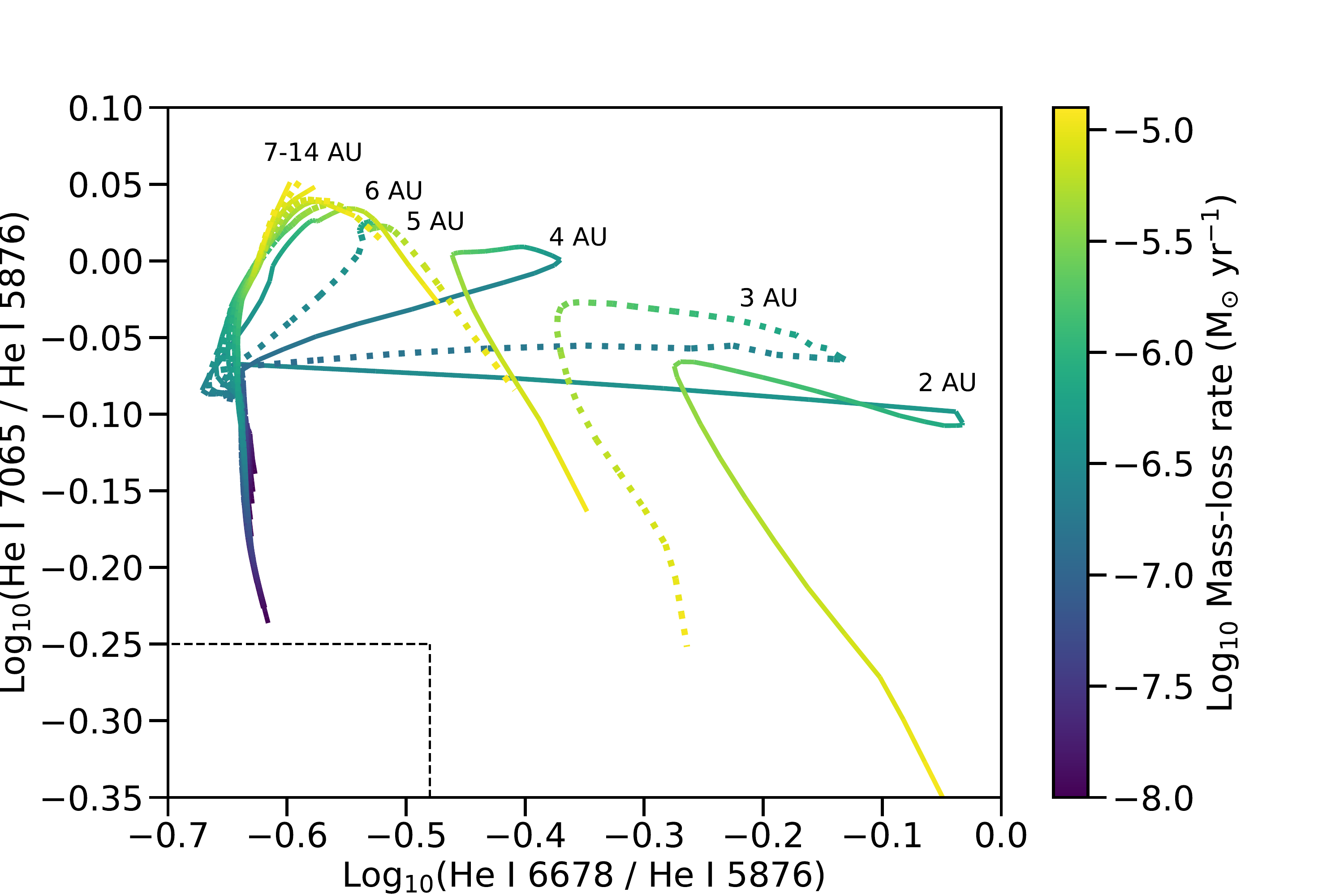}
\caption{The He \textsc{i} line ratio diagram. The X-axis shows the He \textsc{i} 6678/He \textsc{i} 5876 line ratio and the y-axis shows the He \textsc{i} 7065/He \textsc{i} 5876 line ratio. The alternating solid and dashed lines show the line ratios calculated for various orbital separations from 2 AU to 14 AU. The colour of the lines show the mass-loss rate according to the colourbar on the right. The black dashed lines shows the classification criterion proposed by \citet{Ilkiewicz17}: objects above and to the right of this line are classified as symbiotic binaries. }
\label{fig:lineDiagram_HE}
\end{figure}

All of the traditional line ratio diagrams considered above rely on the forbidden emission lines. However, there are many symbiotic binaries with few or no forbidden emission lines in the optical spectrum. For example, the only forbidden line present in LIN 358 is the [Fe \textsc{x}] 6374 \AA\ line \citep{Kuuttila21}. For such sources different classification criteria are needed. 
One promising method is to use the He \textsc{i} lines, which are also bright in LIN 358. 
The He \textsc{i} lines are quite sensitive to the density of the CSM, because they are mainly collisionally excited from the meta-stable $2^3$S level. Thus these lines originate from near the donor star where the density is the highest. 

We show in Figure~\ref{fig:lineDiagram_HE} the He \textsc{i} line ratio diagram with He \textsc{i} 6678/He \textsc{i} 5876 vs. He \textsc{i} 7065/He \textsc{i} 5876 line ratios, but this time for many different orbital separations ranging from 2 AU to 14 AU. Below the steady burning regime all the models are nearly indistinguishable, regardless of the orbital separation and mass-loss rate, and the line ratios stay close to the Case B values (e.g. He \textsc{i} 6678/He \textsc{i} 5876 $\approx$ 0.25). However, at higher mass-loss rates and small separations the ratios deviate significantly from the these values due to the collisional depopulation of the meta-stable $2^3$S to the higher states \citep{Bray00, Osterbrock06}. With separations of 2--4 AU especially the He \textsc{i} 6678/He \textsc{i} 5876 line ratio increases significantly. For separations of $\gtrsim$ 7 AU, the line ratios are nearly independent of the separation. 
Similar behaviour has been previously observed by \citet{Proga94}, who used these line ratios to distinguish between the long period D-type and the shorter period S-type symbiotic binaries.  
Their results show that D-type symbiotic binaries have typically He \textsc{i} 6678/He \textsc{i} 5876 $\sim$ 0.25 while the S-type symbiotic binaries have He \textsc{i} 6678/He \textsc{i} 5876 $\gtrsim$ 0.5. Our results are in agreement and show that the main difference is the binary separation.

\section{Summary}\label{sec:conclusions}

In this work we have studied emission line spectra of symbiotic binaries with photoionization simulations. We have used a 2D simulations method to run simulations with the spectral synthesis code \C, as previously presented in \citet{Kuuttila21}. 
For the given  mass-loss rate of the donor star and the binary separation, we calculate the mass-accretion rate, temperature, and luminosity of the white dwarf in a self-consistent manner.  Using \C\ we solve the emission and ionization structure of the circumstellar medium around the white dwarf and study how the ionization state of the gas and   luminosities of optical emission lines originating from the ionized gas depend on the binary parameters. 

With our simulations we show that the circumstellar medium around symbiotic binaries is mostly ionized (Fig.~\ref{fig:columnDepth}). Especially in the steady nuclear burning regime the CSM is almost fully ionizied, save for a narrow cone towards the donor star, and thus there is no significant obscuration of the WD emission by the dense stellar winds from the donor star. This is important in the context of ionized nebulae around white dwarfs \citep{Rappaport94}, because such nebulae should then exist also around symbiotic binaries, although none have been detected up to date \citep{Remillard95}. This is also related to the single degenerate type Ia supernova progenitor channel, wherein the progenitor is expected to create a long lasting nebula, which can be used to constrain the progenitor channels \citep{2013MNRAS.432.1640W, Graur14, Woods16, Woods17, Woods18, Kuuttila19, Graur19, Farias20}. 

We also present emission line brightness diagrams of several important emission lines and we show how these line luminosities depend on the orbital separation and mass-loss rate of the donor star (Fig.~\ref{fig:grid}). While the behaviour of many emission lines, e.g. Balmer lines, is quite smooth, some emission lines exhibit more varied behaviour (Figs.~\ref{fig:lineEmission_3.0} and \ref{fig:lineEmission_14.0}). Many forbidden lines show a sharp decrease in luminosity at the onset of nuclear burning on the surface of the WD, when the ionization state of the gas increases. Otherwise, while still faint, there is notable emission in the common nebular forbidden emission lines well beyond their critical densities. 
In addition to the common nebular forbidden lines, we have studied the various forbidden iron emission lines, e.g. [Fe \textsc{x}] 6374 and [Fe \textsc{xiv}] 5303 \AA. These lines have been previously detected in many symbiotic binaries and we show that they are excellent indicators of the nuclear burning on the WD surface. These lines are very sensitive to the colour temperature of the WD and thus the emission peaks sharply in the stable nuclear burning regime where the temperature is the highest, while the iron lines with lower ionization state, e.g. [Fe \textsc{vii}] 6087, decrease in brightness in the same regime due to the increased ionization state of the gas (Figs.~\ref{fig:lineEmission_3.0} and \ref{fig:lineEmission_14.0}). 

We have explored the various emission line ratio diagrams used to separate symbiotic binaries from other sources \citep[e.g.][]{Proga94, Gutierrez-Moreno95, Ilkiewicz17}. 
There is a broad consistency between our calculations and the previously derived empirical line ratio diagrams.  
We have shown how the various line ratios change with the orbital separation and mass-loss rate, how this is seen in the context of S- and D-type symbiotic binaries, and how useful the certain line ratio diagrams are in separating symbiotic binaries with e.g. orbital period (Figs.~\ref{fig:lineDiagram_1},~\ref{fig:lineDiagram_2}, and \ref{fig:lineDiagram_HE}). 
Many of these line ratios involve forbidden lines, which may be absent in many symbiotic binaries, making the usage of these diagrams difficult. For these systems, a useful tool is the He \textsc{i} line ratio diagram (Fig.~\ref{fig:lineDiagram_HE}), which has been previously used to distinguish between S- and D-type symbiotic binaries \citep{Proga94}. The helium lines are especially sensitive to the density and thus serve as a clear indicator of the distance between the WD and the donor star.

\section*{Acknowledgements}

The authors would like to thank the referee, Prof. Gary Ferland, for inspiring comments. 
 MG acknowledges partial support by the RSF grant 19--12--00369.

\section*{Data Availability Statement}

The data underlying this article will be shared on reasonable request to the corresponding author.




\bibliographystyle{mnras}
\bibliography{Symbiotic_binaries}


\appendix

\section{1.0 M$_{\odot}$ white dwarf simulations}

\begin{figure}
\centering
\includegraphics[width=0.47\textwidth]{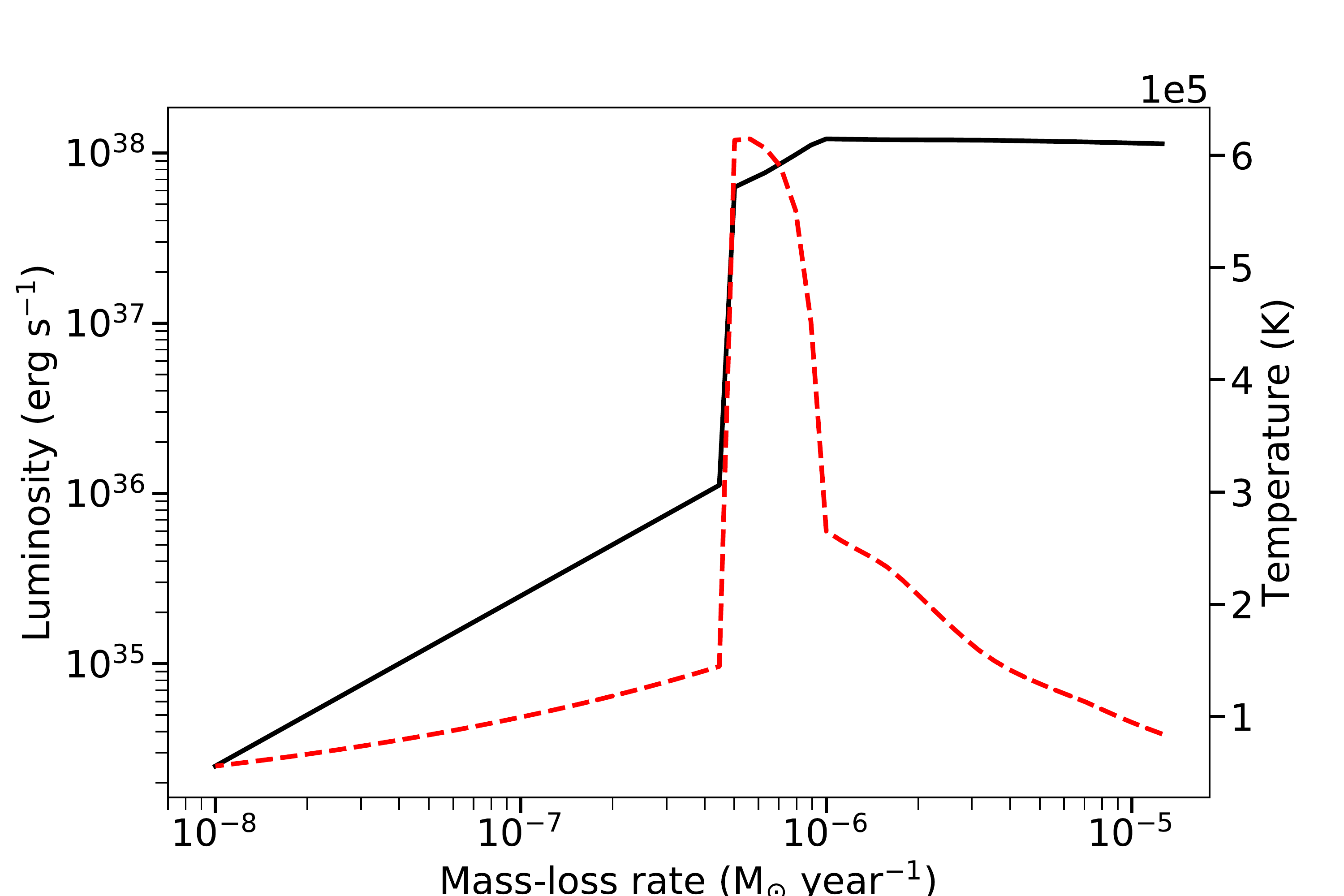}
\includegraphics[width=0.47\textwidth]{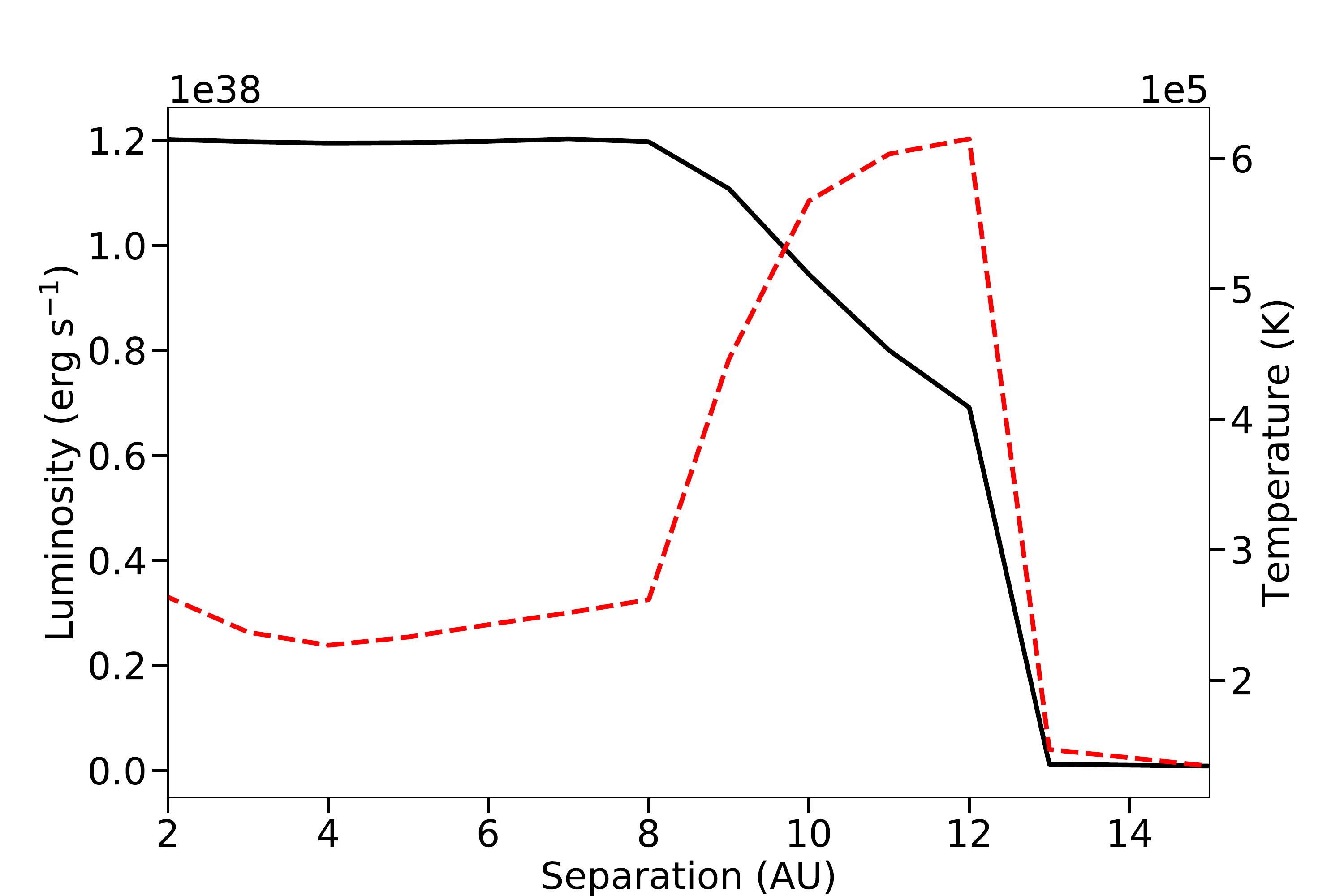}
\caption{The same as Fig.~\ref{fig:temp_lum} but for 1.0 \msun\ WD. 
\textit{Upper panel:} The luminosity and temperature of the WD as a function of the mass-loss rate from the donor star with constant separation of 5 AU. The y-axis on the left shows the luminosity in erg s$^{-1}$ (black line) in logarithmic scale, and the y-axis on the right shows the temperature in K (red dashed line) in linear scale. The insert on the top left shows the luminosity in linear scale in the high mass-loss regime. The sharp rise in both temperature and luminosity at $\sim 2 \times 10^{-7}$ \msun\ yr$^{-1}$ is because of the onset of the nuclear burning on the WD. \newline
\textit{Lower panel:} The luminosity and temperature of the WD as a function of separation with a constant mass-loss rate of $10^{-5.8}$ \msun\ yr$^{-1}$. The y-axis on the left shows the luminosity (black line) in linear scale, and the y-axis on the right shows the temperature (red dashed line) in linear scale. For low mass-loss rates with no steady burning the temperature shown is the maximum temperature of the accretion disk.}
\label{fig:1.0_temp_lum}
\end{figure}

\begin{figure}
\centering
\includegraphics[width=0.47\textwidth]{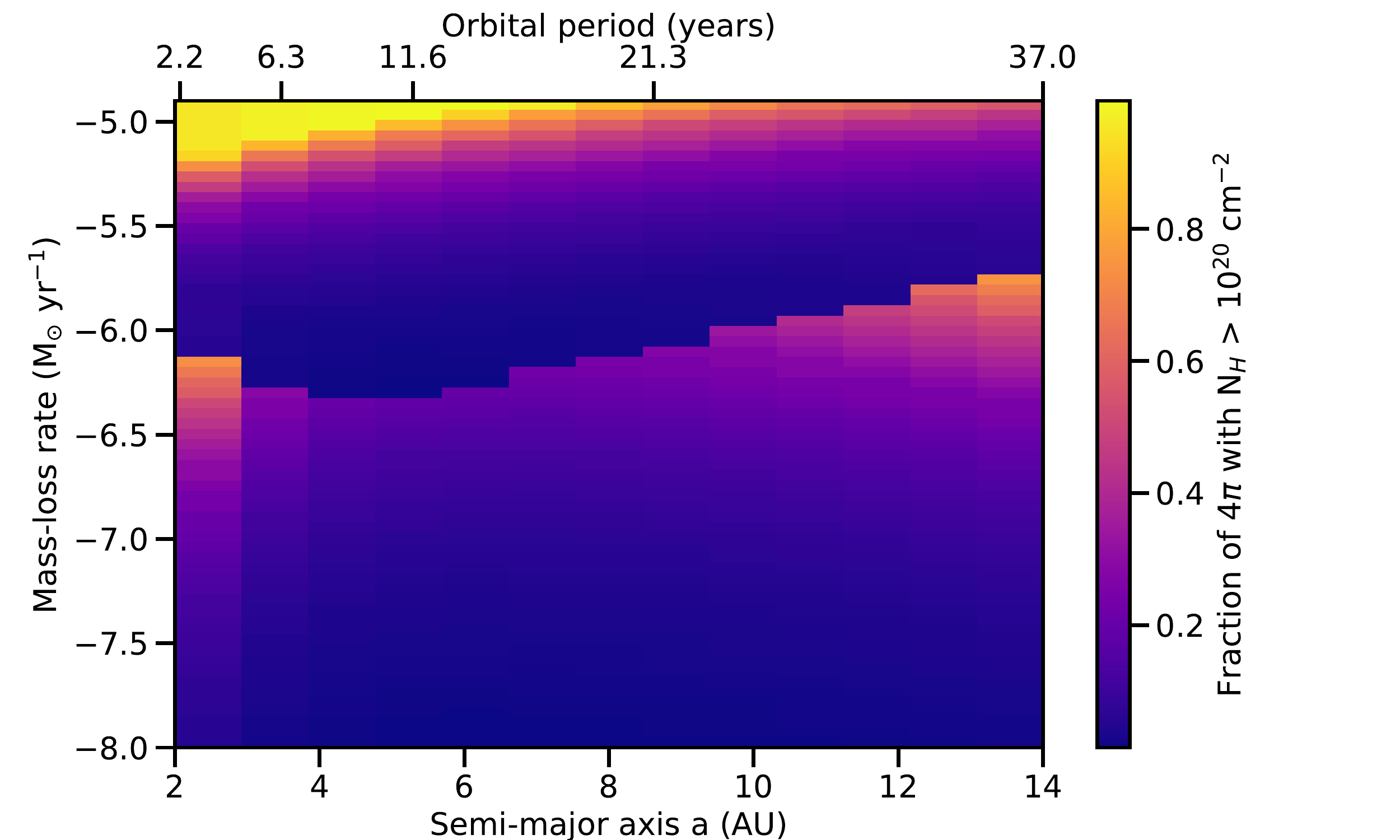}
\caption{The same as Fig.~\ref{fig:columnDepth} but for 1.0 \msun\ WD. 
The neutral fraction of the CSM around the WD. The x-axis shows the orbital separation at the bottom and the Keplerian orbital period on top, and the y-axis shows the mass-loss rate from the donor star. The colour in this figure illustrates the fraction of the solid angle, where N$_{H} > 10^{20}$ cm$^{-2}$, minus the fraction of the sky covered by the donor star. The jump between 10$^{-6.5}$ and 10$^{-6.0}$ \msun\ yr$^{-1}$ is caused by the onset of steady nuclear burning on the WD. }
\label{fig:1.0_columnDepth}
\end{figure}

\begin{figure}
\centering
\includegraphics[width=0.47\textwidth]{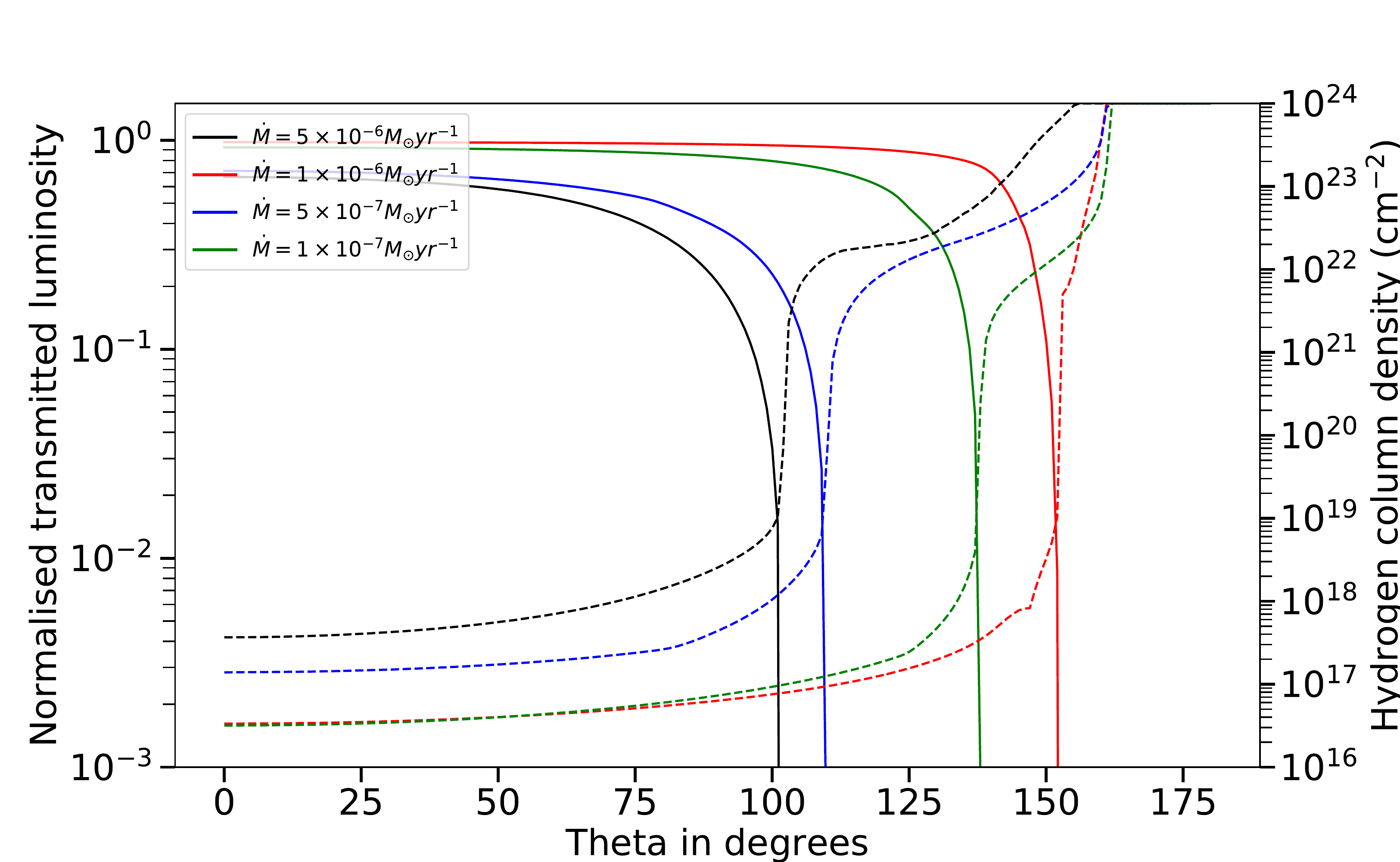}
\caption{The same as Fig.~\ref{fig:attenuation_angle} but for 1.0 \msun\ WD. 
The transmitted luminosity of the WD (solid lines) and the CSM neutral hydrogen column density (dashed lines) as a function of the angle $\theta$. The y-axis on the left side show the transmitted luminosity normalised to the initial WD luminosity, and the y-axis on the right show the column density in units of cm$^{-2}$. The black, red, blue, and green lines show the results for separation of 3 AU and mass-loss rates of $5 \times 10^{-6}$, $1 \times 10^{-6}$, $5 \times 10^{-7}$, and $1 \times 10^{-7}$ \msun\ yr$^{-1}$, respectively. }
\label{fig:1.0_attenuation_angle}
\end{figure}

\begin{figure}
\centering
\includegraphics[width=0.47\textwidth]{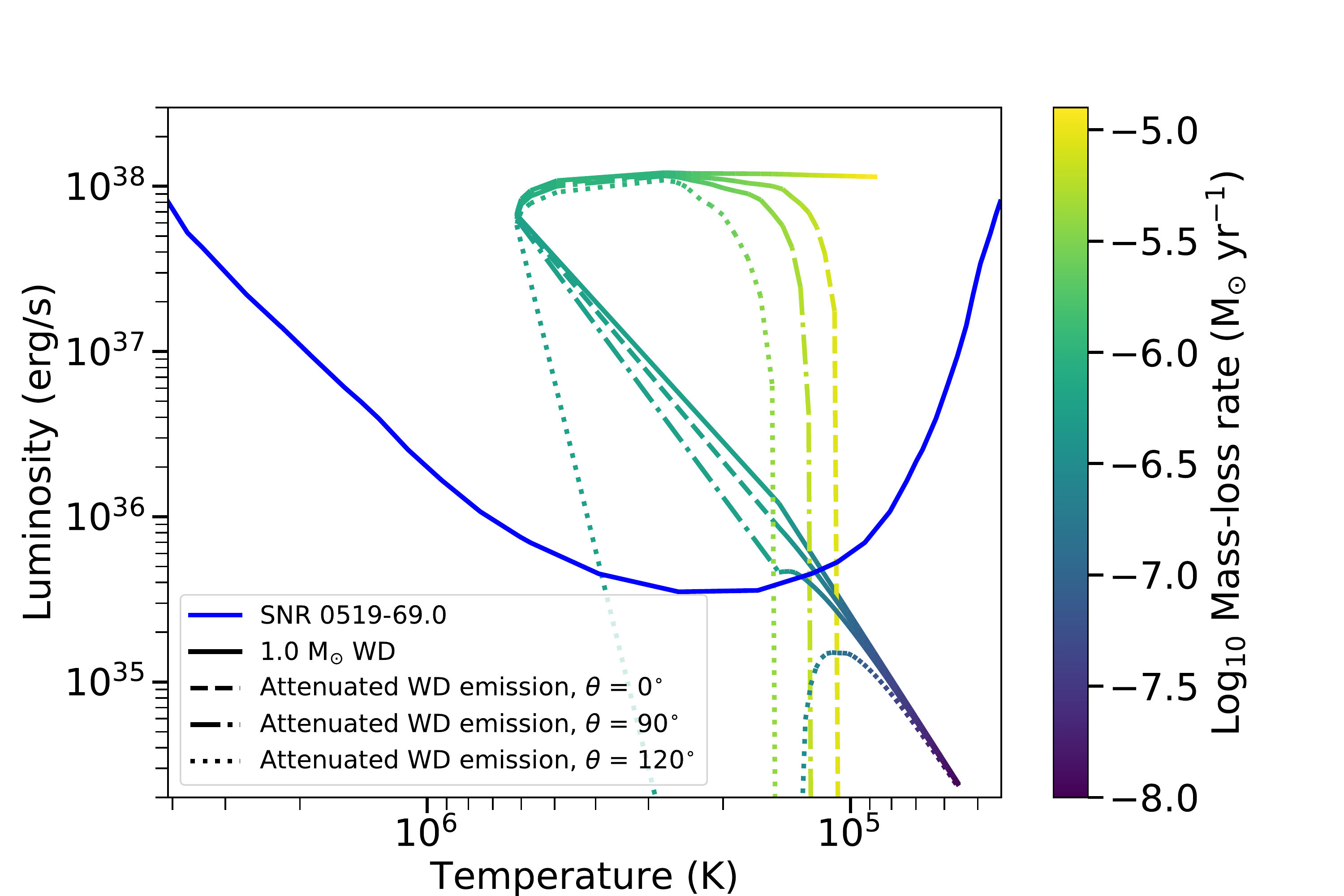}
\caption{The same as Fig.~\ref{fig:lum_temp_attenuation} but for 1.0 \msun\ WD. 
The temperature and luminosity of the WD as a function of mass-loss rate. The solid coloured line shows the initial WD temperature and luminosity, while the dashed, dash-dotted, and dotted lines show the attenuated emission for angles 0$^{\circ}$, 90$^{\circ}$, and 120$^{\circ}$, respectively. The mass of the WD is 1.0 \msun\ and the orbital separation is 3 AU.
The solid blue line shows the upper limit on the temperature and luminosity of the progenitor of type Ia SNR 0519-69.0 from \citet{Kuuttila19}. }
\label{fig:1.0_lum_temp_attenuation}
\end{figure}

\begin{figure*}
\centering
\includegraphics[width=\textwidth]{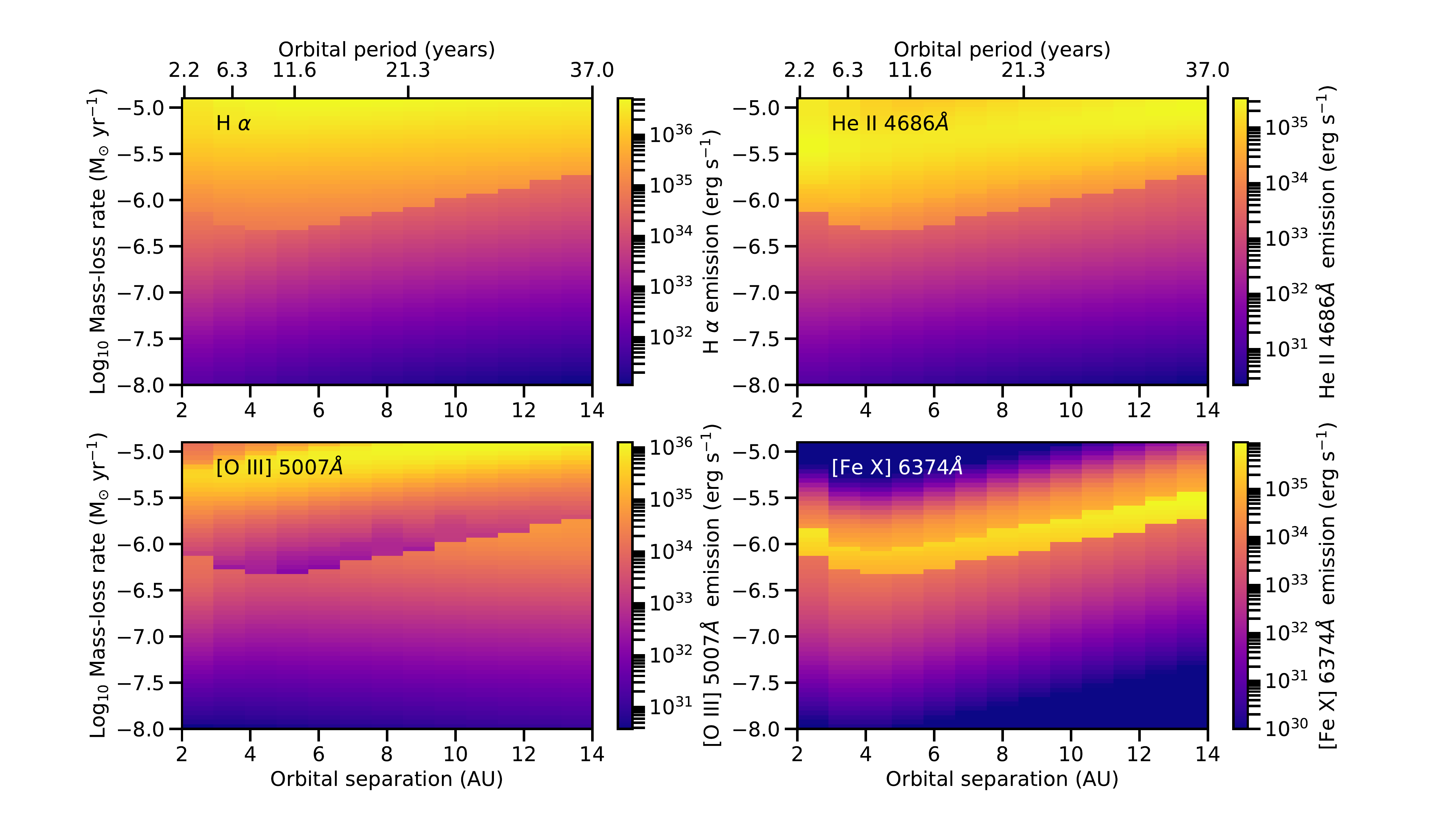}
\caption{The same as Fig.~\ref{fig:grid} but for 1.0 \msun\ WD. 
The line luminosity (in erg s$^{-1}$) for four different emission lines as predicted by the \C\ simulations. The x-axis shows the orbital separation $r_c$ in AUs and the Keplerian orbital period in years for a WD mass 1.0 \msun. The colour shows the line luminosity for H$\alpha$ (top left), He \textsc{ii} 4686 (top right), [O \textsc{iii}] 5007 (bottom left), and [Fe \textsc{x}] 6374 \AA\ (bottom right). }
\label{fig:1.0_grid}
\end{figure*}

\begin{figure*}
\centering
\includegraphics[width=\textwidth]{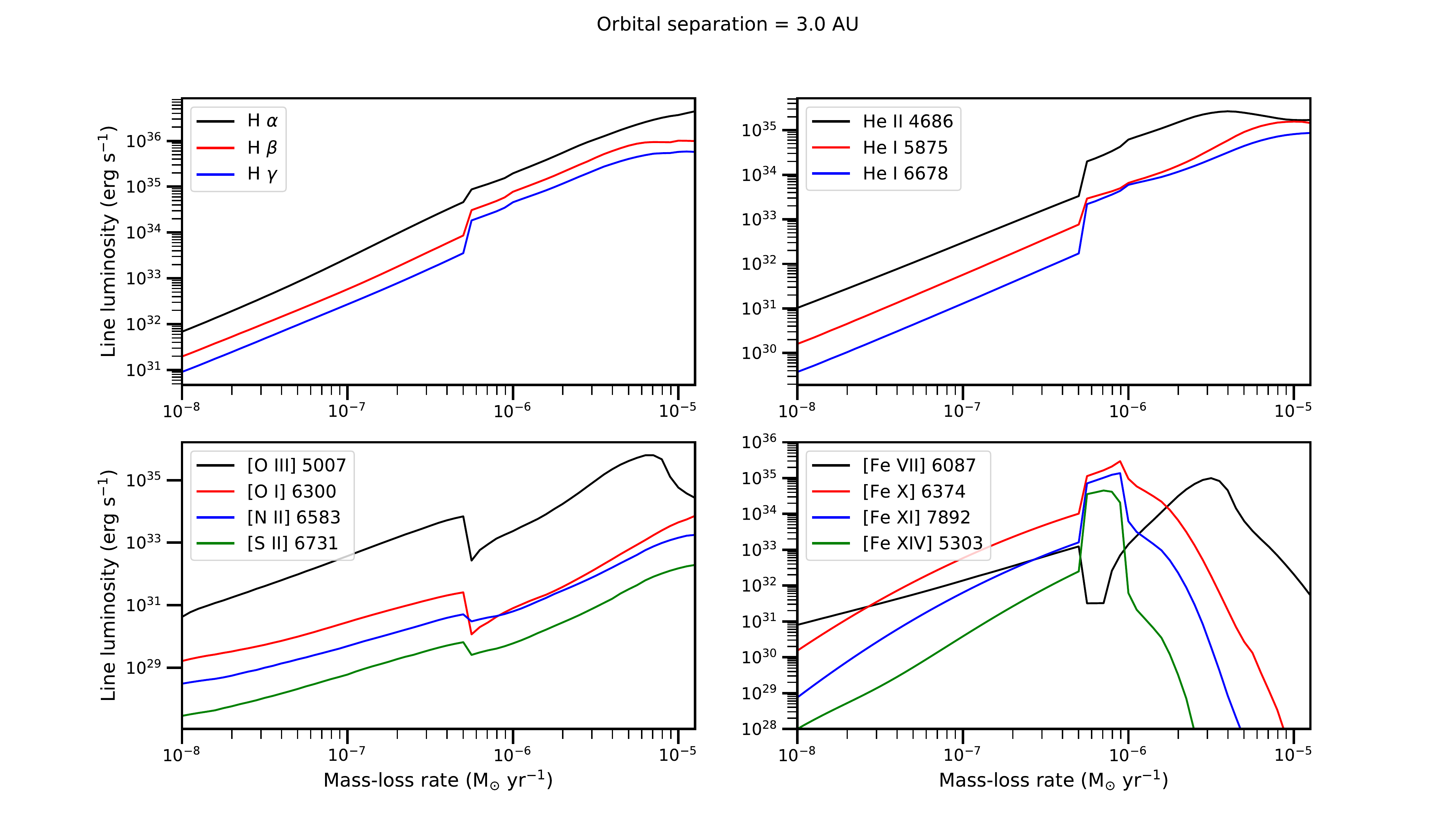}
\caption{The same as Fig.~\ref{fig:lineEmission_3.0} but for 1.0 \msun\ WD. 
The line luminosity (in erg s$^{-1}$) as a function of the mass-loss rate for various emission lines. The orbital separation is set to 3 AU. The top left panel shows the line luminosity for H$\alpha$ (black line), H$\beta$ (red line), and H$\gamma$ (blue line). The top right shows the He \textsc{ii} 4686 (black), He \textsc{i} 5875 (red), and He \textsc{i} 6678 \AA\ lines. The bottom left shows the common forbidden [O \textsc{iii}] 5007 (black), [O \textsc{i}] 6300 (red), [N \textsc{ii}] 6583 (blue), and [S \textsc{ii}] 6731 \AA\ (green) emission lines. The bottom right panel shows various iron lines: [Fe \textsc{vii}] 6087 (black), [Fe \textsc{x}] 6374 (red), [Fe \textsc{xi}] 7892 (blue), and [Fe \textsc{xiv}] 5303 \AA\ (green).}
\label{fig:1.0_lineEmission_3.0}
\end{figure*}

\begin{figure*}
\centering
\includegraphics[width=\textwidth]{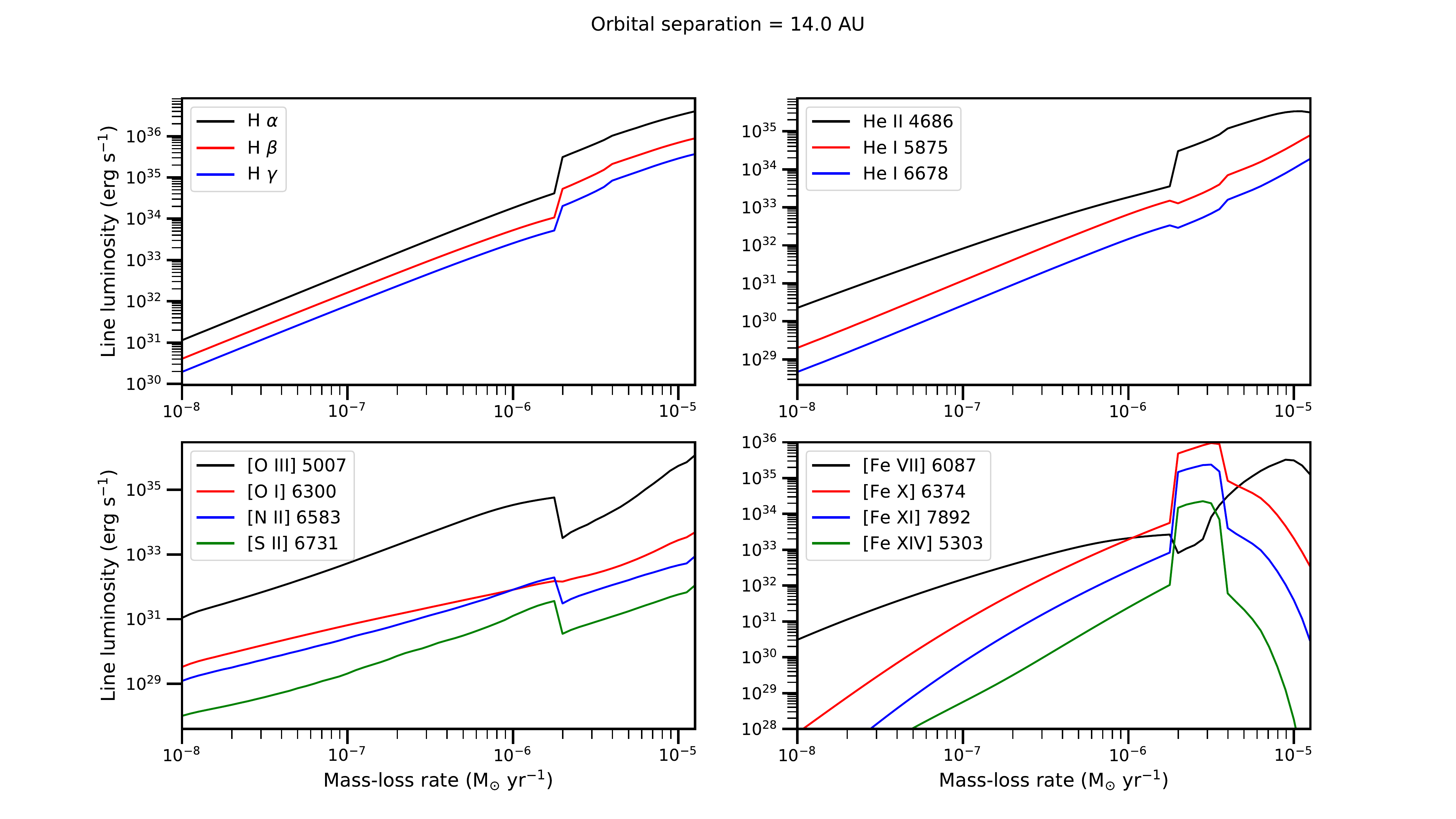}
\caption{The same as Fig.~\ref{fig:1.0_lineEmission_3.0} but for separation of 14 AU. }
\label{fig:1.0_lineEmission_14.0}
\end{figure*}

\begin{figure}
\centering
\includegraphics[width=0.5\textwidth]{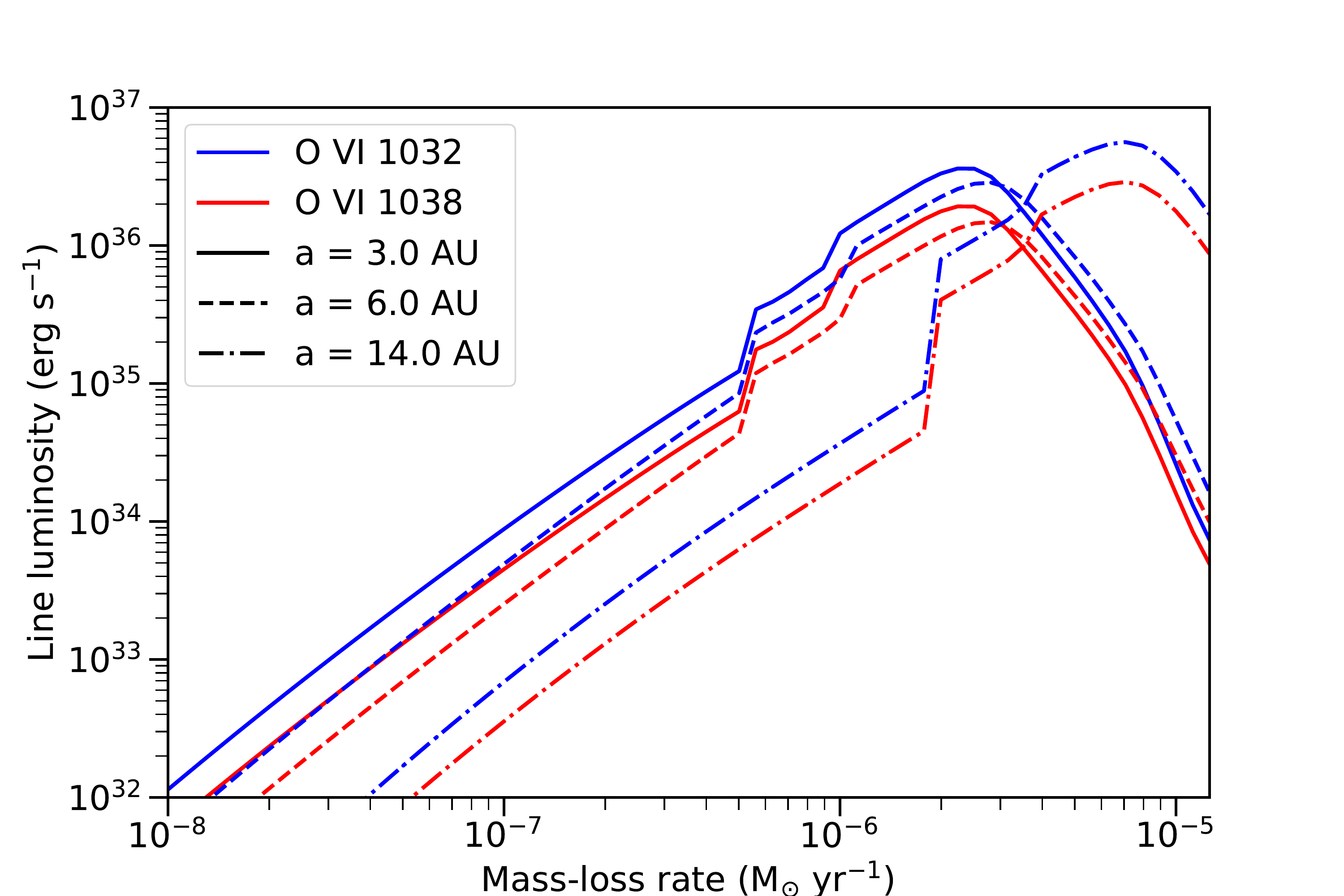}
\caption{The same as Fig.~\ref{fig:UVlines} but for 1.0 \msun\ WD.
The line luminosity of O \textsc{vi} 1032 \AA\ (blue), 1038 \AA\ (red) emission lines as a function of mass-loss rate for orbital separations of 3.0 (solid lines), 6.0 (dashed lines), and 14.0 AU (dash-dotted lines). }
\label{fig:1.0_UVlines}
\end{figure}

\begin{figure}
\centering
\includegraphics[width=0.47\textwidth]{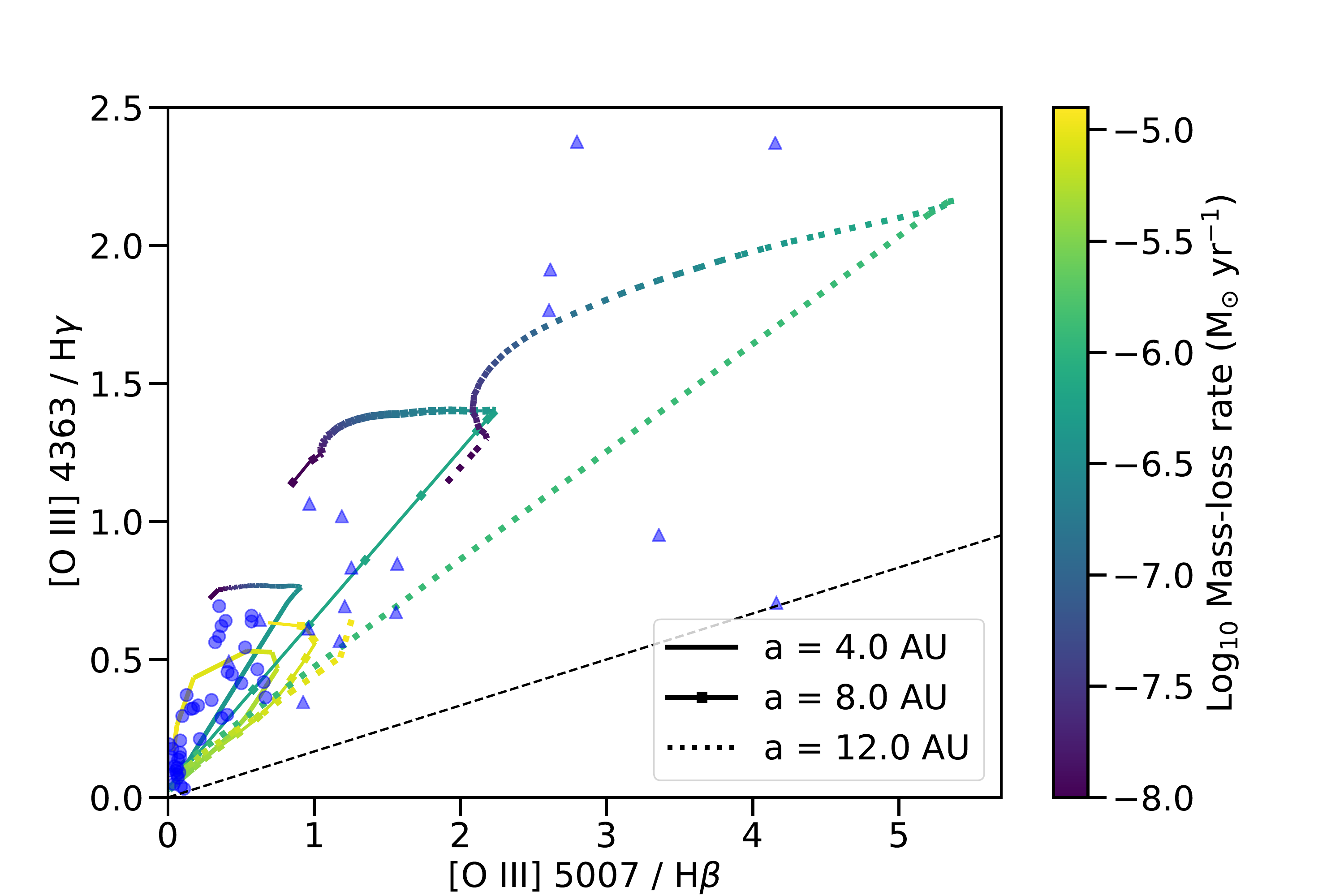}
\caption{The same as Fig.~\ref{fig:lineDiagram_1} but for 1.0 \msun\ WD. 
[O \textsc{iii}] 5007/H$\beta$ vs. [O \textsc{iii}] 4363/H$\gamma$ line ratio diagram of \citet{Gutierrez-Moreno95}. The solid, solid-dotted, and the dotted lines show the line ratios for 4, 8, and 12 AU separations, respectively. The colour of the lines show the mass-loss rate according to the colourbar on the right. The blue circles show the S-type and the blue triangles show the D-type symbiotic binaries. The black dashed lines shows the classification criterion of \citet{Gutierrez-Moreno95}: objects above this line are classified as symbiotic binaries. }
\label{fig:1.0_lineDiagram_1}
\end{figure}

\begin{figure*}
\centering
\includegraphics[width=0.47\textwidth]{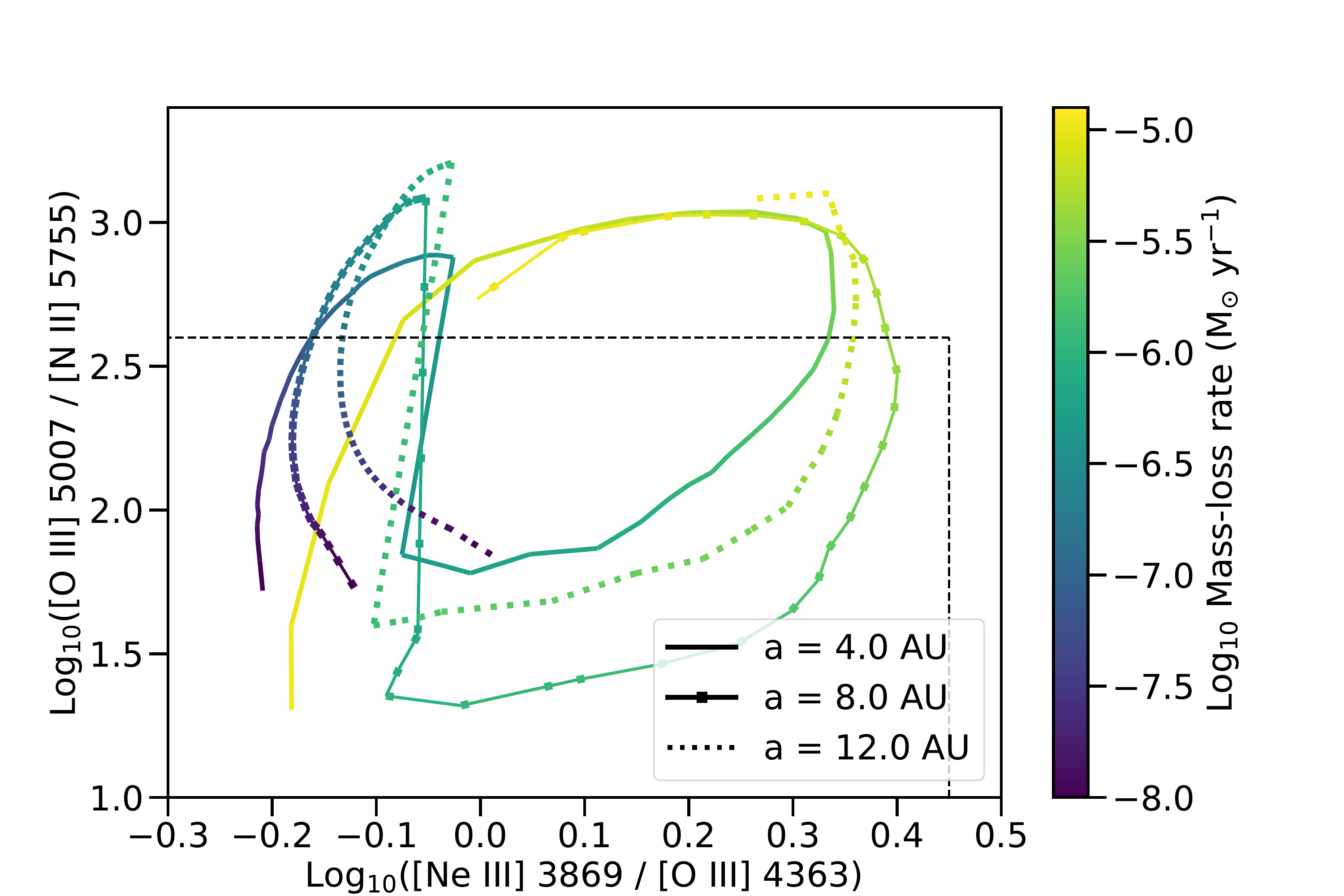}
\includegraphics[width=0.47\textwidth]{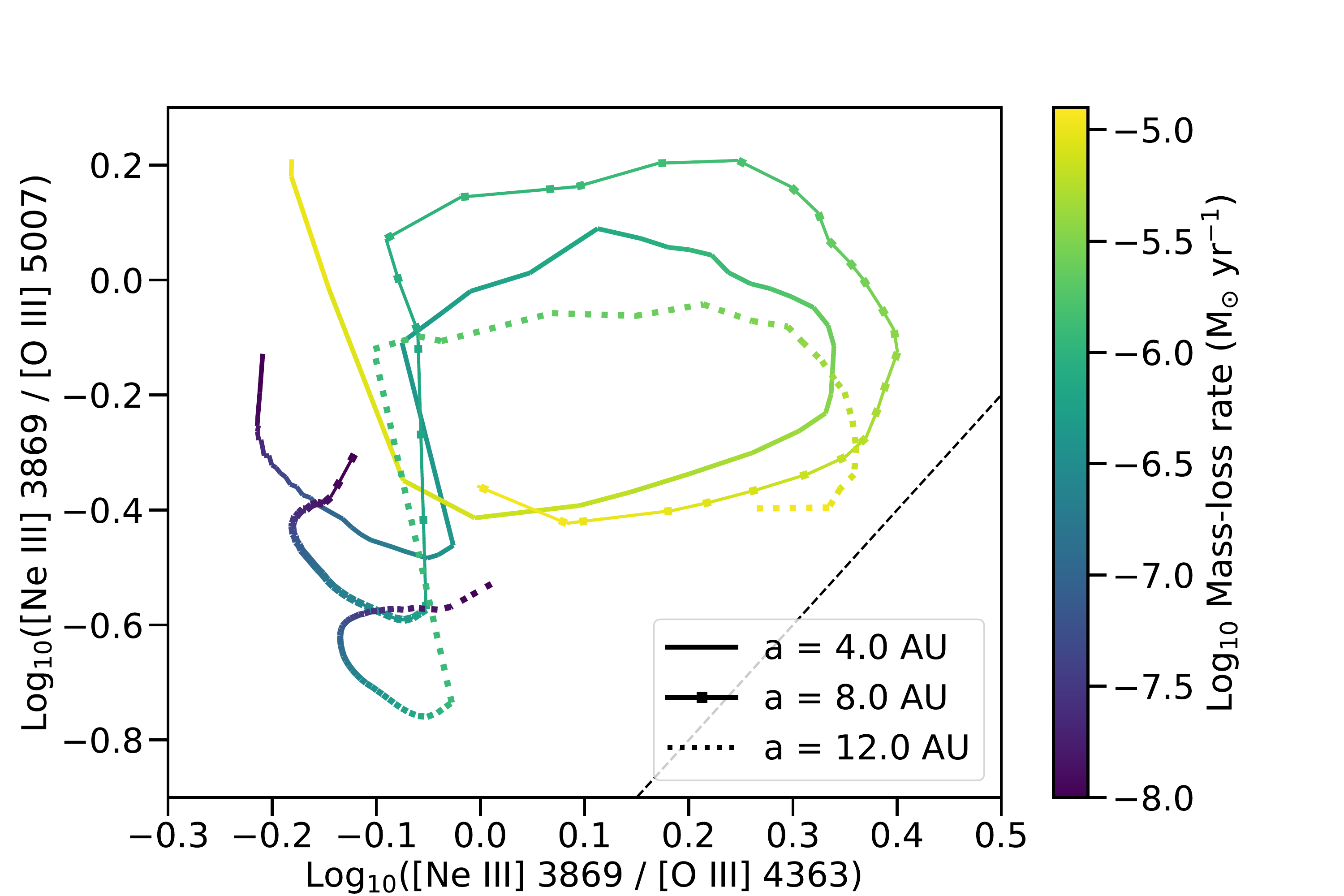}
\caption{The same as Fig.~\ref{fig:lineDiagram_2} but for 1.0 \msun\ WD. 
Two different line ratio diagrams of \citet{Ilkiewicz17}. 
On the left is shown [Ne \textsc{iii}] 3869/[O \textsc{iii}] 4363 vs. [O \textsc{iii}] 5007/[N \textsc{ii}] 5755 and on the right is shown [Ne \textsc{iii}] 3869/[O \textsc{iii}] 4363 vs. [Ne \textsc{iii}] 3869/[O \textsc{iii}] 5007 line ratios. 
The solid, solid-dotted, and the dotted lines show the line ratios for 4, 8, and 12 AU separations, respectively. The colour of the lines show the mass-loss rate according to the colourbar on the right. The black dashed lines show the the criteria for symbiotic binaries proposed by \citet{Ilkiewicz17}.}
\label{fig:1.0_lineDiagram_2}
\end{figure*}

\begin{figure}
\centering
\includegraphics[width=0.47\textwidth]{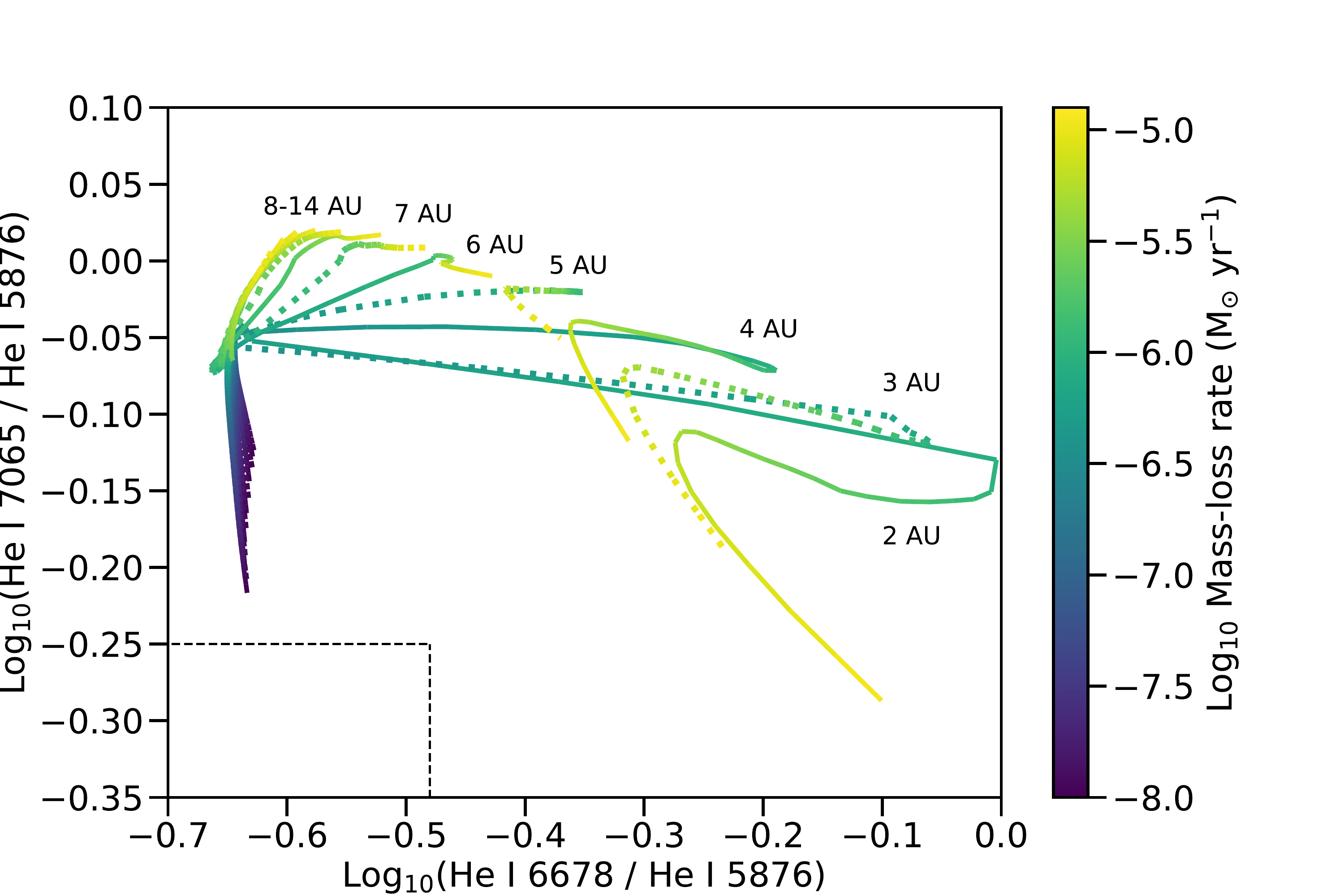}
\caption{The same as Fig.~\ref{fig:lineDiagram_HE} but for 1.0 \msun\ WD. 
The He \textsc{i} line ratio diagram. The X-axis shows the He \textsc{i} 6678/He \textsc{i} 5876 line ratio and the y-axis shows the He \textsc{i} 7065/He \textsc{i} 5876 line ratio. The alternating solid and dashed lines show the line ratios calculated for various orbital separations from 2 AU to 14 AU. The colour of the lines show the mass-loss rate according to the colourbar on the right. The black dashed lines shows the classification criterion proposed by \citet{Ilkiewicz17}: objects above and to the right of this line are classified as symbiotic binaries. }
\label{fig:1.0_lineDiagram_HE}
\end{figure}


\bsp	
\label{lastpage}
\end{document}